 \documentclass[manuscript, screen]{acmart}
\AtBeginDocument{%
  }

\UseRawInputEncoding
\usepackage{etoolbox}

\usepackage{algorithmic}
\usepackage{array}
\usepackage[caption=false,font=normalsize,labelfont=sf,textfont=sf]{subfig}
\usepackage{textcomp}
\usepackage{stfloats}
\usepackage{url}
\usepackage{ragged2e} 
\usepackage{xcolor}
\usepackage{graphicx}  % For rotating text
\usepackage{arydshln}  % For dashed lines
\usepackage{multirow}  % Already included for multirow
\usepackage{amsmath}
\usepackage{siunitx}
\usepackage{tikz}
\usetikzlibrary{positioning, math, fit, backgrounds}
\usepackage{adjustbox}
\usepackage{balance}
\usepackage{svg}
\usepackage{tabularx, soul,xcolor}
\usepackage{threeparttable}
\usepackage{booktabs} % For better table styling
\usepackage{caption} % To customize caption
\usepackage{pifont} 
\DeclareMathOperator*{\argmax}{argmax}
\DeclareMathOperator*{\argmin}{argmin}
\usepackage{comment}
\usepackage{colortbl}

\usepackage[normalem]{ulem}
\useunder{\uline}{\ul}{}
\usepackage{mathtools}
\usepackage{booktabs} 
\usepackage{xspace}
\usepackage{bbding}
\usepackage{pifont}

\newcommand{\xmark}{\ding{56}}
\usepackage{acronym}

\newacro{ml}[ML]{machine learning}
\newacro{dl}[DL]{deep learning}
\newacro{ai}[AI]{artificial intelligence}
\newacro{nlp}[NLP]{natural language processing}
\newacro{dnn}[DNN]{deep neural network}
\newacro{vit}[ViT]{vision transformer}
\newacro{dynn}[DyNN]{dynamic neural network}
\newacro{qoe}[QoE]{quality of experience}
\newacro{uap}[UAP]{universal perturbation attack}
\newacro{cnn}[CNN]{convolutional neural network}
\newacro{ae}[AE]{adversarial example}
\newacro{lbfgs}[L-BFGS]{limited memory broyden-fletcher-goldfarb-shanno}
\newacro{fgm}[FGM]{fast gradient method}
\newacro{bim}[BIM]{basic iterative method}
\newacro{pgd}[PGD]{projected gradient descent}
\newacro{zoo}[ZOO]{zeroth order optimization}
\newacro{relu}[ReLU]{rectified linear unit}
\newacro{as}[AS]{attack success}
\newacro{cas}[CAS]{constrained attack success}
\newacro{gpu}[GPU]{graphics processing unit}
\newacro{tpu}[TPU]{tensor processing unit}
\newacro{asic}[ASIC]{application
specific integrated circuit}
\newacro{sgd}[SGD]{stochastic gradient descent}
\newacro{dos}[DoS]{denial-of-service}
\newacro{ga}[GA]{genetic algorithm}
\newacro{adnn}[AdNN]{adaptive neural network}
\newacro{nms}[NMS]{non-maximum suppression}
\newacro{iou}[IoU]{intersection over union}
\newacro{fpga}[FPGA]{field-programmable gate array}
\newacro{pdn}[PDN]{power distribution network}
\newacro{ode}[ODE]{ordinary differential equation}
\newacro{itp}[ITP]{input transferability percentage}
\newacro{etp}[ETP]{effectiveness transferability percentage}
\newacro{flops}[FLOPs]{floating-point operations per second}
\newacro{dct}[DCT]{discrete cosine transform}
\newacro{snn}[SNN]{spiking neural network}
\newacro{ann}[ANN]{artificial neural network}
\newacro{nicgm}[NICGM]{neural image caption generation model}
\newacro{sos}[SOS]{special token}
\newacro{eos}[EOS]{end of sequence}
\newacro{vlm}[VLM]{vision-language model}
\newacro{nmt}[NMT]{neural machine translation}
\newacro{ce}[CE]{cross entropy}
\newacro{mlaas}[MLaaS]{machine learning as a service}
\newacro{cv}[CV]{computer vision}
\newacro{deit}[DeiT]{data-efficient image transformer}
\newacro{ereba}[EREBA]{energy robustness using estimator-based approach}
\newacro{mlaas}[MLaaS]{Machine Learning as a Service}
\newacro{sota}[SOTA]{state-of-the-art}

\begin{document}

\title{Energy-Latency Attacks: A New Adversarial Threat to Deep Learning}

\author{Hanene F. Z. Brachemi Meftah}
\affiliation{%
  \institution{Univ. Rennes, INSA Rennes, CNRS, IETR - UMR 6164}
  \city{Rennes}
  \country{France}
}
\email{hanene.brachemi@insa-rennes.fr}
\orcid{0009-0008-1399-4506}

\author{Wassim Hamidouche}
\affiliation{%
  \institution{KU 6G Research Center, Khalifa University}
  \city{Abu Dhabi}
  \country{UAE}
}
\email{whamidouche@gmail.com}
\orcid{0000-0002-0143-1756}
\author{Sid Ahmed Fezza}
\affiliation{\institution{National Higher School of Telecommunications and ICT}
  \city{Oran}
  \country{Algeria}}
\email{sfezza@ensttic.dz}
\orcid{0000-0001-6453-8588}
\author{Olivier D\'eforges}
\affiliation{%
  \institution{Univ. Rennes, INSA Rennes, CNRS, IETR - UMR 6164}
  \city{Rennes}
  \country{France}
}
\email{olivier.deforges@insa-rennes.fr}
\orcid{0000-0003-0750-0959}

\renewcommand{\shortauthors}{Brachemi Meftah et al.}
%\begin{abstract}
%The growing computational needs of \acp{dnn} have raised concerns about their energy consumption. To address these challenges, energy-efficient hardware and \acp{dnn} have become essential for sustainable deployment of \ac{ai}. However, these efficiency-focused designs have shown vulnerabilities that attackers can exploit to increase latency and energy usage. This paper provides an overview of current research on energy-latency attacks, categorizing them using the established taxonomy for adversarial attacks. We also explore the different metrics used to evaluate the performance of these attacks, compare different attack strategies, analyze existing defenses, and highlight current challenges and potential areas for future research in this developing field. The GitHub page for this work can be accessed at: \href{https://github.com/hbrachemi/Survey_energy_attacks/}{https://github.com/hbrachemi/Survey\_energy\_attacks/}
%[\textit{hidden for anonymous purposes}].}
%\end{abstract}
\begin{abstract}
The growing computational demand for \acp{dnn} has raised concerns about their energy consumption and carbon footprint, particularly as the size and complexity of the models continue to increase. To address these challenges, energy-efficient hardware and custom accelerators have become essential. Additionally, adaptable \acp{dnn} are being developed to dynamically balance performance and efficiency. The use of these strategies became more common to enable sustainable \acs{ai} deployment. However, these efficiency-focused designs may also introduce vulnerabilities, as attackers can potentially exploit them to increase latency and energy usage by triggering their worst-case-performance scenarios. This new type of attack, called energy-latency attacks, has recently gained significant research attention, focusing on the vulnerability of \acp{dnn} to this emerging attack paradigm, which can trigger \ac{dos} attacks. This paper provides a comprehensive overview of current research on energy-latency attacks, categorizing them using the established taxonomy for traditional adversarial attacks. We explore different metrics used to measure the success of these attacks and provide an analysis and comparison of existing attack strategies. We also analyze existing defense mechanisms and highlight current challenges and potential areas for future research in this developing field. The GitHub page for this work can be accessed at \href{https://github.com/hbrachemi/Survey_energy_attacks/}{https://github.com/hbrachemi/Survey\_energy\_attacks/}.
\end{abstract}
\begin{CCSXML}
<ccs2012>
   <concept>
       <concept_id>10003752.10003809.10010047.10010051</concept_id>
       <concept_desc>Theory of computation~Adversary models</concept_desc>
       <concept_significance>500</concept_significance>
       </concept>
   <concept>
       <concept_id>10002978.10003006.10011610</concept_id>
       <concept_desc>Security and privacy~Denial-of-service attacks</concept_desc>
       <concept_significance>300</concept_significance>
       </concept>
 </ccs2012>
\end{CCSXML}

\ccsdesc[500]{Theory of computation~Adversary models}
\ccsdesc[300]{Security and privacy~Denial-of-service attacks}
\keywords{Adversarial attacks, Energy-latency attacks,  Deep neural networks}


\maketitle
\acresetall
\section{Introduction}
In the past decade, \acp{dnn} have attracted unprecedented attention across diverse fields, including \ac{nlp}, \ac{cv}, healthcare, autonomous driving, and finance. This surge in interest is largely due to their ability to learn complex patterns and representations from massive datasets, leading to significant performance improvements in various tasks. Notably, the recent trend of scaling these models, increasing their size, data, and computational resources, has unlocked even greater potential, pushing the boundaries of what \acp{dnn} can achieve~\cite{10.5555/3600270.3602446}.
The widespread adoption of \acp{dnn}, while transformative, has introduced substantial computational and energy burdens~\cite{strubell2019energy}.  The increase in \ac{dnn} applications coincides with a heightened awareness of energy consumption and a growing emphasis on green technologies. In particular, Strubell {\it et al.}~\cite{strubell2019energy} projected in 2019 that a 50\% reduction in carbon emissions was critical in the current decade to avoid a resurgence of natural disasters. However, the increasing complexity of training \ac{dnn} models exacerbates this environmental challenge. As an illustration, training a BERT architecture~\cite{devlin2018bert} on a V100x64 GPU for 79 hours can generate 652.3 kg of $\text{CO}_2$ emissions, compared to the 11.8~kg emitted when training a simpler transformer architecture~\cite{vaswani2017attention} on a P100x8 GPU for 12 hours. Furthermore, the substantial energy requirements of these models pose a challenge, even with the increasing availability of renewable energy sources. Current limitations in renewable energy production and storage infrastructure limit their ability to fully offset the demands of \ac{dl}~\cite{strubell2019energy}.

Consequently, companies face the complex challenge of optimizing model performance while simultaneously managing latency constraints, computational resources, and energy costs. This requires a two-pronged approach involving both software and hardware innovations. On the hardware side, specialized accelerators are used, while algorithmic optimizations rely on techniques such as exploiting data sparsity and locality, implementing power and clock management, utilizing parallel processing, and adopting reduced-precision computing. The synergistic application of these hardware and software strategies aims to minimize energy consumption and reduce processing latency, thereby achieving significant computational gains without compromising model accuracy~\cite{yin2022vit,meng2022adavit,dettmers2022gpt3,chen2019eyeriss,han2016eie}.

%idea 2 : problematic (adversaries)
Given the need to optimize \ac{dl} effectiveness, it is crucial to address the rigorous scrutiny that \acp{dnn} have faced regarding their robustness and security in adversarial settings. Over the past decade, considerable research has focused on evaluating the integrity~\cite{goodfellow2014explaining,madry2017towards,carlini2017towards,moosavi2015deepfool,chen2017zoo,chen2020hopskipjumpattack,andriushchenko2020square,liu2017neural,cheng2021deep,gu2017badnets,liu2020reflection,nguyen2021wanet,LLuCY18,10.1145/3465397} and confidentiality~\cite{carlini2022membership,shokri2017membership,leino2020stolen,sablayrolles2019white,carlini2019secret,dinur2003revealing,nguyen2023active,chen2022amplifying} of \acp{dnn}. However, the cybersecurity triad of confidentiality, integrity, and availability was not fully considered in the context of \acp{dnn} until Shumailov {\it et al.}~\cite{shumailov2021sponge} introduced a new attack vector targeting availability, known as a \textit{sponge attack}.  This attack, also termed an energy-latency attack, is designed to compromise the availability of \ac{dl} platforms by substantially increasing their energy consumption and/or response latency, potentially leading to a \ac{dos} scenario. Following this pioneering work, several studies~\cite{chen2022amplifying,cina2022energy,pan2022gradauto,chen2023dark} have extended this concept, developing attacks that increase the energy consumption or latency of the model for purposes beyond \ac{dos}. The consequences of such attacks are far-reaching, including increased operational energy costs, inflated service fees, and decreased \ac{qoe} for end-users. For example, Shumailov {\it et al.}~\cite{shumailov2021sponge} demonstrated a case study where they increased the response time of Microsoft Azure's translator by a factor of 6000x. Other research~\cite{wang2023energy,cina2022energy} have shown that this type of attack can effectively drain the battery life of a victim user's mobile device. Furthermore, some energy-latency attacks exert a more insidious impact, gradually increasing energy costs over extended periods. Although seemingly negligible in the short term, these actions can accumulate, leading to substantial economic and environmental repercussions. The significance of addressing these attacks is further underscored by their ability to neutralize the effects of engineered optimizations on the system. By forcing the system to operate in a default, i.e.,~non-optimized state, these attacks effectively decrease the energy efficiency gains achieved through dedicated research and development, ultimately leading to increased costs and wasted resources. As \ac{ai} systems become increasingly integrated into critical sectors such as industry, healthcare, finance, and autonomous driving, ensuring their security and reliability is paramount.  Addressing vulnerabilities like those exploited by sponge attacks is essential to mitigate potential safety risks, financial losses, system failures, and degraded performance in these vital applications.

The emergence of energy-latency attacks has prompted the research community to investigate the vulnerabilities of \ac{dl} models. These attacks are often categorized based on existing attack typologies, as illustrated in Fig.~\ref{fig:schema}. Key criteria include the attack stage (inference vs. training), knowledge access (white-box vs. black-box), and control level (partial vs. full).  However, effective energy-latency attacks require a detailed understanding of the prediction pipeline. Consequently, many energy-latency attacks are meticulously crafted to target specific pipeline subprocesses~\cite{hong2020panda,haque2020ilfo,chen2023dark,boutros2020neighbors,liu2023slowlidar}. This new attack paradigm has led to numerous studies exploring the feasibility of such attacks across diverse scenarios and architectures~\cite{hong2020panda,haque2020ilfo,chen2023dark,boutros2020neighbors,liu2023slowlidar}. While surveys exist for accuracy-based~\cite{gao2020backdoor,aldahdooh2022adversarial,10.1145/3465397,10.1145/3485133,10.1145/3398394,10.1145/3547330,10.1145/3691625,10.1145/3636551,10.1145/3594869,10.1145/3702638} and privacy-based attacks~\cite{rodriguez2023survey,rigaki2023survey,jegorova2022survey}, a comprehensive analysis of energy-latency attacks is lacking. Therefore, the aim of this paper is to fill this gap by providing a systematic categorization, summary, and analysis of current research in this rapidly evolving field. This review will serve as a valuable resource for the research community, clarifying the prerequisites, applications, and limitations of existing attacks. Ultimately, this will facilitate the development of effective defenses and guide future research directions.

The remainder of this paper is structured as follows. Section~\ref{sec:prelim} provides an overview of traditional adversarial attacks and their relationship to energy-latency attacks. Section~\ref{sec:works} presents a summary and taxonomy of existing energy-latency attacks, detailing their application context and formulation. Subsequently, Section~\ref{sec:metrics} presents an exhaustive list of metrics used to evaluate the effectiveness of various existing sponge attacks. A comparative analysis of these attacks is then provided in Section~\ref{sec:analysis}, followed by a review of existing defenses, current challenges and future perspectives within this research field in Section~\ref{sec:openchal}. Finally, Section~\ref{sec:conclusion} concludes and summarizes this paper.

\section{Preliminaries}
\label{sec:prelim}
\subsection{Brief Overview of Adversarial Attacks}
\begin{figure}[t!]
    \centering
    \includegraphics[width=1\linewidth]{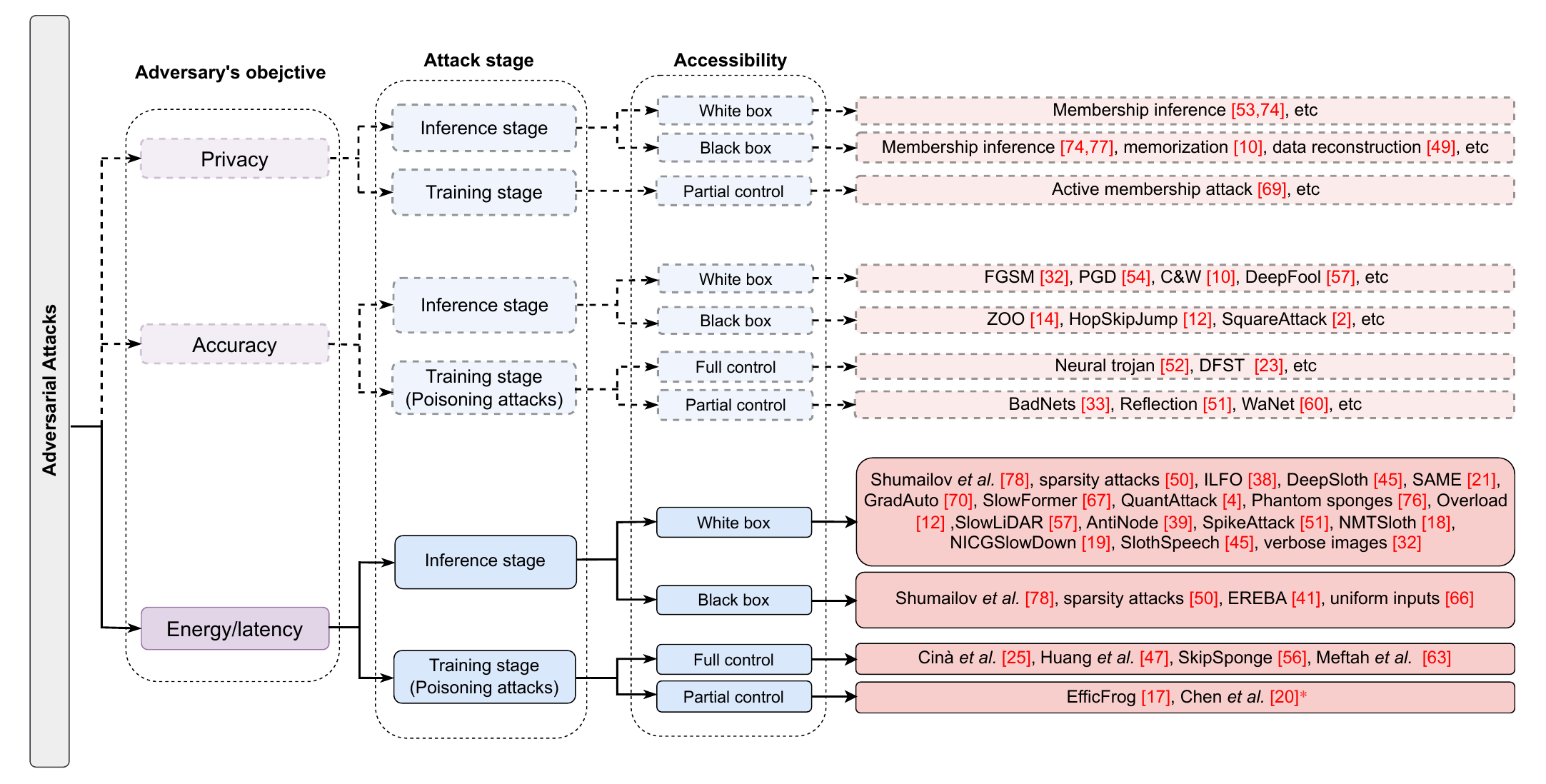}
    \caption{Categorization of adversarial attacks based on objective, control, and stage. \textcolor{red}{*}Chen \textit{et al.}~\cite{chen2023stealthy}  have a distinct objective of increasing training time and cost, unlike mainstream methodologies in the literature that focus on increasing inference time and energy. }
    \label{fig:schema}
\end{figure}

\begin{figure}[t!]
    \centering
    \includegraphics[width=1.07\linewidth]{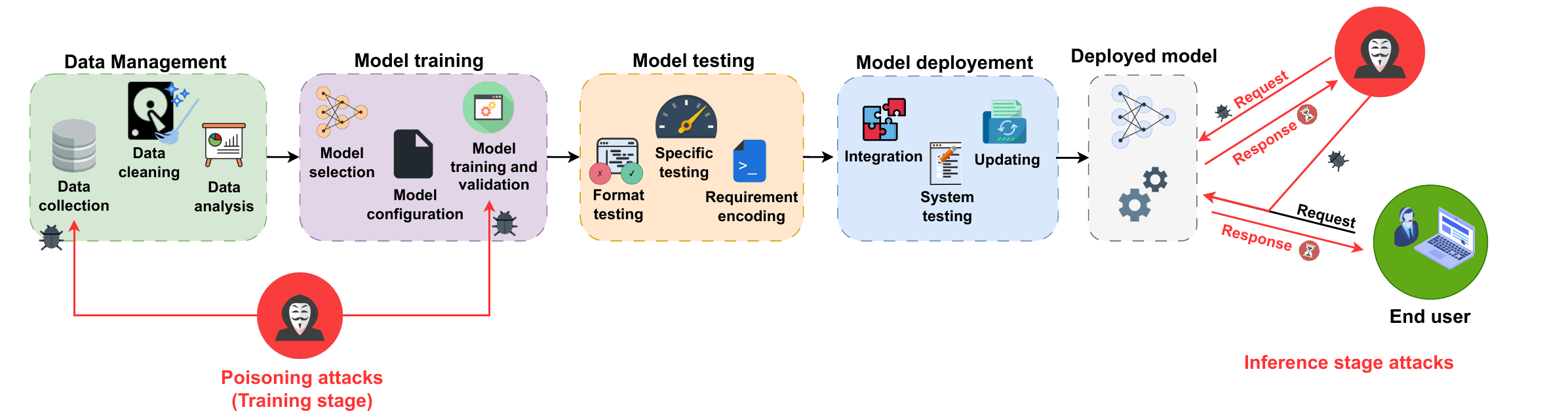}
    \caption{Different attack scenarios throughout the model lifecycle.}
    \label{fig:model_life_cycle}
\end{figure}

Adversarial attacks involve intentionally manipulating input data to exploit vulnerabilities in \ac{dl} models. Szegedy {\it et al.}~\cite{szegedy2013intriguing} first revealed these vulnerabilities, sparking further research into the weaknesses of \ac{dl} models~\cite{goodfellow2014explaining,madry2017towards,carlini2017towards,moosavi2015deepfool,chen2017zoo,chen2020hopskipjumpattack,andriushchenko2020square,liu2017neural,cheng2021deep,gu2017badnets,liu2020reflection,nguyen2021wanet,carlini2022membership,shokri2017membership,leino2020stolen,sablayrolles2019white,carlini2019secret,dinur2003revealing,nguyen2023active,chen2022amplifying,LLuCY18}. Consequently, the research community has dedicated significant effort to develop robust defense mechanisms. However, this is an ongoing challenge due to the continuous evolution of increasingly sophisticated adversarial attacks~\cite{wang2023adversarial}.  Adversarial attacks can be categorized along multiple dimensions, as shown in Fig.~\ref{fig:schema}. One approach classifies attacks based on the adversary's objective, distinguishing between attacks that target accuracy, privacy, or energy-latency. Another categorization considers the attacker's access to the victim model, differentiating between white-box, gray-box, and black-box attacks. Attacks can also be classified by the attacker's level of control during the training process (full or partial). Furthermore, a methodological distinction can be made between inference stage attacks, which manipulate model inputs, and training stage (or poisoning) attacks, which target model weights. This latter classification, illustrated in Fig.~\ref{fig:model_life_cycle}, is often termed \textit{attack stage-based}, as inference stage attacks typically occur during the inference phase, while poisoning attacks occur during training or prior to deployment. For a more in-depth understanding, readers can refer to comprehensive surveys on adversarial attacks~\cite{gao2020backdoor,aldahdooh2022adversarial,rodriguez2023survey,rigaki2023survey,jegorova2022survey}.
\subsection{Towards Energy-latency Attacks}
Energy-latency attacks represent a novel category of adversarial attacks, defined by their unique objectives. Despite their recent emergence, these attacks can be categorized using the same taxonomy as accuracy- and privacy-based attacks, taking into account the attacker's access and the stage of the attack. They also share similarities with accuracy-based attacks in their method of generating adversarial examples, typically by optimizing a specific loss function with respect to the input to modify the model's behavior. As a result, many existing accuracy-based attacks~\cite{goodfellow2014explaining, madry2017towards, carlini2017towards} can be adapted for energy-latency attacks by altering the optimization objective. These new objectives can directly target the increase of latency and energy consumption or focus on model-specific aspects such as activation sparsity, the number of activated blocks, or the minimization of pruned tokens. Despite the varied objectives, these attacks share a common goal: to increase the computational load on the target model. 

Unlike traditional accuracy-based attacks~\cite{goodfellow2014explaining, madry2017towards, carlini2017towards, moosavi2015deepfool, chen2017zoo, chen2020hopskipjumpattack, andriushchenko2020square, liu2017neural, cheng2021deep, gu2017badnets, liu2020reflection, nguyen2021wanet}, which primarily focus on manipulating model outputs or features, energy-latency attacks~\cite{pan2022gradauto, hong2020panda, haque2020ilfo, haque2023antinode, chen2022nmtsloth} take a broader approach. These attacks require a more comprehensive understanding and exploitation of the target model's internal mechanisms. Consequently, they are highly dependent on the specific architecture, processing pipeline, and execution environment.
\section{Energy-latency Attack Strategies}
%\setlabel{EXISTING SPONGE ATTACKS STRATEGIES}
\label{sec:works}
In this section, we provide a comprehensive overview of existing energy-latency attacks, as classified in Fig.~\ref{fig:schema}. We delve into their requirements, formulation, and application context to provide a thorough understanding of these attacks. Table~\ref{tab:existing_attacks} summarizes the proposed energy-latency attacks in the literature while Table~\ref{tab:notations} presents a recap of the notations used throughout the paper.
\begin{table*}[t!]
\centering
\caption{Summary of the proposed energy-latency attacks in the literature. $^\star$ symbol refers to white-box attacks that can be transferred to other architectures in a black-box setting. Chen \textit{et al.}~\cite{chen2023stealthy}$^\oplus$ have a distinct objective of increasing training time and cost, unlike mainstream methodologies in the literature that focus on increasing inference time and energy.}
\label{tab:existing_attacks}
\resizebox{\textwidth}{!}{%
\begin{tabular}{cl|c|>{\raggedright}p{3cm}|c|>{\raggedright}p{3.5cm}|l|c|c}
\toprule
\rowcolor[HTML]{DAE8FC} 
Attack stage &
  Attack &
  Venue &
  Target &
  Application &
  Attack objective &
  Used datasets &
  Attack setting &
  Open source \\ \midrule
 &
   &
   &
  Transformers &
  NLP &
  Increase hidden activations' dimensions &
  \begin{tabular}[c]{@{}l@{}}SuperGLUE benchmark, WMT translation\end{tabular} &
  White box &
   \\ \cmidrule{4-8}
 &
   &
   &
  CNNs &
  CV &
  Minimize activations' sparsity &
  ImageNet &
  White box &
   \\ \cmidrule{4-8}
 &
  \multirow{-5}{*}{Shumailov et \textit{al.}~\cite{shumailov2021sponge}} &
  \multirow{-5}{*}{EuroS\&P (2021)} &
  Transformers &
  NLP &
  Increase model's latency &
  \begin{tabular}[c]{@{}l@{}}SuperGLUE benchmark, WMT translation\end{tabular} &
  Black box &
  \multirow{-5}{*}{\CheckmarkBold} \\ \cmidrule{2-9} 
 &
  Sparsity attacks~\cite{krithivasan2020sparsity} &
    \parbox{4cm}{Transactions on Computer-Aided Design of Integrated Circuits and Systems (2020)} &
  CNNs &
  CV &
  Minimize activations' sparsity &
  MNIST, CIFAR10, ImageNet &
  White box$^\star$ &
  \xmark \\ \cmidrule{2-9} 
 &
  ILFO~\cite{haque2020ilfo} &
  CVPR (2020) &
  Depth-dynamic AdNNs &
  CV &
  Maximize the number of activated layers &
  CIFAR10, ImageNet &
  White box &
  \xmark \\ \cmidrule{2-9} 
 &
  DeepSloth~\cite{hong2020panda} &
  ICLR (2021) &
  Multi-exit CNNs &
  CV &
  Bypass early-exits &
  \begin{tabular}[c]{@{}l@{}}CIFAR10, Tiny ImageNet, CIFAR100\end{tabular} &
  White box &
  \CheckmarkBold \\ \cmidrule{2-9} 
 &
  GradAuto~\cite{pan2022gradauto} &
  ECCV (2022) &
  Depth- and width-dynamic AdNNs &
  CV &
  Maximize the number of activated layers and channels &
  CIFAR10, ImageNet &
  White box &
  \CheckmarkBold \\ \cmidrule{2-9} 
 &
  SpikeAttack~\cite{krithivasan2022efficiency} &
  ACM-DAC (2022) &
  SNNs &
  CV &
  Maximize the number of memory accesses &
  CIFAR10, ImageNet &
  White box &
  \xmark \\ \cmidrule{2-9} 
 &
  NMTSloth~\cite{chen2022nmtsloth} &
  ESEC/FSE (2022) &
  Decoder-based NMT systems &
  NLP &
  Maximize the number of decoder calls &
  MultiUN &
  White box &
  \CheckmarkBold \\ \cmidrule{2-9} 
 &
  NICGSlowDown~\cite{chen2022nicgslowdown} &
  CVPR (2022) &
  Decoder-based NICG models &
  CV &
  Maximize the number of decoder calls &
  Flickr8k, MS COCO &
  White box &
  \CheckmarkBold \\ \cmidrule{2-9} 
 &
  EREBA~\cite{haque2022ereba} &
  ICSE-SEIP (2022) &
  AdNNs &
  CV &
  Maximize the energy estimation &
  CIFAR10, CIFAR100 &
  Black box &
  \xmark \\ \cmidrule{2-9} 
 &
  SAME~\cite{chen2023dynamic} &
  ACL (2023) &
  Multi-exit transformers &
  NLP &
  Bypass early-exits &
  GLUE benchmark &
  White box &
  \CheckmarkBold \\ \cmidrule{2-9} 
 &
  SlowFormer~\cite{navaneet2023slowformer} &
  CVPR (2024) &
  Efficient ViTs &
  CV &
  Recover pruned tokens by the efficient ViT &
  ImageNet, CIFAR10 &
  White box &
  \CheckmarkBold \\ \cmidrule{2-9} 
 &
  QuantAttack~\cite{baras2023quantattack} &
  WACV (2025) &
  ViTs, DeiT &
  CV &
  Maximize used precision for feature vectors and activations &
  \begin{tabular}[c]{@{}l@{}}ImageNet, COCO, FLEURS\end{tabular} &
  White box$^\star$ &
  \xmark \\ \cmidrule{2-9} 
 &
  Phantom Sponge~\cite{shapira2023phantom} &
  WACV (2023) &
  Object detection &
  CV &
  Minimize the number of candidates discarded by NMS &
  \begin{tabular}[c]{@{}l@{}}BDD, MTSD, LISA\end{tabular} &
  White box$^\star$ &
  \CheckmarkBold \\ \cmidrule{2-9} 
 &
  Overload~\cite{chen2023overload} &
  CVPR (2023) &
  Object detection &
  CV &
  Minimize the number of candidates discarded by NMS &
  MS COCO &
  White box &
  \xmark \\ \cmidrule{2-9} 
 &
  SlowLiDAR~\cite{liu2023slowlidar} &
  CVPR (2023) &
  Object detection &
  CV &
  Minimize the number of candidates discarded by NMS &
  KITTI &
  White box$^\star$ &
  \CheckmarkBold \\ \cmidrule{2-9} 
 &
  AntiNODE~\cite{haque2023antinode} &
  CVPR (2023) &
  ODE &
  CV &
  Minimize the step size &
  CIFAR10, MNIST &
  White box$^\star$ &
  \CheckmarkBold \\ \cmidrule{2-9} 
 &
  SlothSpeech~\cite{haque2023slothspeech}&
  Interspeech (2023) &
  Decoder-based speech recognition models &
  NLP &
  Maximize the number of decoder calls &
  \begin{tabular}[c]{@{}l@{}}LibriSpeech, OpenSLR, VCTK\end{tabular} &
  White box &
  \CheckmarkBold \\ \cmidrule{2-9} 
 &
  Verbose images~\cite{gao2024inducing} &
  ICLR (2024) &
  Decoder-based VLMs &
  CV &
  Maximize the number of decoder calls &
  MS COCO, ImageNet &
  White box$^\star$ &
  \CheckmarkBold \\ \cmidrule{2-9} 
\multirow{-40}{*}[9ex]{\rotatebox{90}{Inference stage}} &
  Uniform inputs~\cite{muller2024impact} &
  SPW (2024) &
  CNNs &
  CV &
  Minimize activations' sparsity &
  ImageNet &
  Black box &
  \CheckmarkBold \\ \midrule
 &
  Cina et \textit{al.}~\cite{cina2022energy} &
  arXiv (2022) &
  CNNs &
  CV &
  Minimize activations' sparsity &
  \begin{tabular}[c]{@{}l@{}}CIFAR10, GTSRB, CelebA\end{tabular} &
  Full control &
  \CheckmarkBold \\ \cmidrule{2-9} 
 &
  EfficFrog~\cite{chen2023dark} &
  CVPR (2023) &
  Multi-exit CNNs &
  CV &
  Bypass early-exits &
  CIFAR10, ImageNet &
  Partial control &
  \CheckmarkBold \\ \cmidrule{2-9} 
 &
  Chen et \textit{al.}~\cite{chen2023stealthy}$^\oplus$ &
  \parbox{3cm}{Journal of Signal Processing Systems (2023)} &
  CNNs &
  CV &
  Increase training time &
  CIFAR10, SVHN &
  Partial control &
  \xmark \\ \cmidrule{2-9} 
 &
  SkipSponge~\cite{lintelo2024spongenet} &
  arXiv (2024) &
  CNNs, GANs &
  CV &
  Minimize activations' sparsity &
  \begin{tabular}[c]{@{}l@{}}MNIST, CIFAR10, GTSRB, Tiny ImageNet\end{tabular} &
  Full control &
  \CheckmarkBold \\ \cmidrule{2-9} 
\multirow{-9}{*}[1ex]{\rotatebox{90}{\centering Training stage}
%\newline\rotatebox{90}{( Poisoning}\newline\rotatebox{90}{  attacks )}
} &
  Huang \textit{et al.}~\cite{huang2024sponge} &
  IEEE Access (2024) &
  Multi-exit CNNs &
  CV &
  Bypass early-exits &
  \begin{tabular}[c]{@{}l@{}}CIFAR100, GTSRB, STL10\end{tabular} &
  Full control &
  \xmark \\ \cmidrule{2-9} 
&
  Meftah \textit{et al.}~\cite{meftah2025energy} &
  ICASSP (2025) &
  CNNs &
  CV &
  Minimize activations' sparsity for triggers&
  \begin{tabular}[c]{@{}l@{}} CIFAR10, Tiny ImageNet\end{tabular} &
  Full control &
  \CheckmarkBold
  \\ \midrule
\end{tabular}%
}
\end{table*}
Although initially designed to cause \ac{dos} on \ac{nlp} platforms~\cite{shumailov2021sponge}, the implications of energy-latency attacks extend far beyond their original scope.  Numerous similar attacks have been developed, each carrying distinct environmental and economic impacts. In practice, these attacks not only amplify latency and energy consumption but also impose higher energy costs on the victim, degrade the \ac{qoe} for end-users, and can potentially undermine the platform's operational integrity. Crafting an effective energy-latency attack requires careful consideration of three key variables in the optimization problem: the attacker's objective (e.g., increasing latency, escalating energy costs, draining user batteries, or inducing \ac{dos}), the attacker's knowledge of and access to the target \ac{dnn} (white-box, gray-box, or black-box), and the attack stage (inference or training). These three variables will form the basis for our discussion of energy-latency attacks in this section.

\begin{table}[t!]
\centering
\caption{List of notations.}
\label{tab:notations}
\resizebox{0.85\textwidth}{!}{
\begin{tabular}{l|l}
\toprule
\rowcolor[HTML]{DAE8FC}
Notation            & Description  \\ \midrule
$\mathbf{x}_{clean}$ & Clean input  \\ \midrule
$\mathbf{x}_{sp}$    & Adversarial example of $\mathbf{x}_{clean}$ \\ \midrule
$\hat{\mathbf{x}}$ & Preprocessed input \\ \midrule
$X$ & Set of clean images  \\ \midrule
$y$ & Ground-truth label of the clean input \\ \midrule
$y_{target}$         & Target label \\ \midrule
$\mathcal{M}$        & Prediction function of the targeted model \\ \midrule
$\delta$             & Perturbation signal introduced to $\mathbf{x}_{clean}$ to form $\mathbf{x}_{sp}$      \\ \midrule
$\oplus$             & \begin{tabular}[c]{@{}l@{}} Operator used to introduce the perturbation into clean text inputs  \end{tabular}    \\ \midrule
$\epsilon$           & Perturbation budget \\  \midrule
$| . |$              & Cardinal of a set \\   \midrule
$\lVert . \rVert_p$  & $\ell_p$-norm \\       \midrule
$\lambda$          &  Lagrangian multiplier \\ \midrule
$\mathcal{L}$      & Loss function \\ \midrule
$\mathcal{T}$        & Preprocessing function \\ \midrule
$\mathcal{T}^{-1}$   & Inverse function of $\mathcal{T}$ \\ \midrule
$\mathcal{A}(\mathbf{x})$ &
  \begin{tabular}[c]{@{}l@{}} Set corresponding to layers' activations within the model during the inference of the input $\mathbf{x}$\end{tabular} \\ \midrule
$a^{(l)} \in \mathcal{A}(\mathbf{x})$ &
  \begin{tabular}[c]{@{}l@{}} Activation of the $l$-th layer during the inference of the input   
  $\mathbf{x}$\end{tabular} \\ \midrule
$\mathcal{P}$        & Set of poisoned samples  \\ \midrule
$\mathcal{N}$        & Normal distribution\\ \midrule
$\mathcal{U}$        &  Uniform distribution \\ \midrule
$\mathcal{M}_i$ &
  \begin{tabular}[c]{@{}l@{}}$i$-th branch of the model $\mathcal{M}$, enabling early exit of the inference process if some constraints are verified\end{tabular} \\ \midrule
$G_l$                & \begin{tabular}[c]{@{}l@{}}The $l$-th layer's gate, determining if the model's $l$-th layer should be activated or discarded\end{tabular} \\ \midrule
$\mathcal{H}$        & Step size of the \acs{ode} solver.  \\ \midrule
$V_i(t)$             & $i$-th membrane potential of the \acs{snn}  at the time step $t$                                 \\ \midrule
$O_i(t)$             &  $i$-th neuron's output of the \acs{snn} at the time step $t$  \\ \midrule
$\text{Loop}_{\mathcal{M}}(\mathbf{x})$ &
Number of decoder calls required for the inference of $\mathbf{x}$ by the model $\mathcal{M}$
  \\ \bottomrule
\end{tabular}
}
\end{table}
\subsection{Inference Stage Attacks} 
Sponge inference stage attacks encompass the meticulous crafting of an input designed to maximize the energy consumption of the target (victim) model. There are two categories of inference stage attacks, depending on the availability of information regarding the targeted model: white-box and black-box attacks.
\subsubsection{\textbf{White-box}}
In a white-box setting, the adversary possesses complete knowledge about their target model, including its architecture and weights for \acs{dnn} models. This level of transparency empowers the attacker to craft a more effective attack strategy that exploits the model's vulnerabilities based on the available information. Consequently, the success of the attacker heavily depends on a comprehensive understanding of the prediction pipeline's functionality. The attacker must strategically formulate the attack by integrating specific criteria into the optimization process and carefully selecting which aspects to target. Numerous energy-latency attacks have been proposed in the white-box setting~\cite{shumailov2021sponge,krithivasan2020sparsity,liu2023slowlidar,navaneet2023slowformer,pan2022gradauto}, each meticulously designed for precise and predefined environments or targeting very specific \ac{dl} models. These include, but are not limited to, attacks tailored for hardware optimizers, input-\acp{adnn}, \ac{nlp} models, and object detection models. In this subsection, we discuss inference stage energy-latency white-box attacks documented in the literature. We categorize these attacks into further subsections based on the architectures they target. Each sub-subsection provides an overview of the architecture's functionality and the basic knowledge required to understand the attack formulation, followed by an explanation of the various proposed attacks within that context.

\paragraph{Sponge attacks on regular \acs{dnn}s}
%\textbf{Background.} 
In their seminal work, Shumailov \textit{et al.}~\cite{shumailov2021sponge} highlight a gap between the average energy consumption of a model and its energy consumption under worst-case conditions. They attribute this to the variability of energy consumed during an inference pass, which depends on two primary factors: the total number of arithmetic operations performed for the input inference and the number of memory accesses. This observation suggests that increasing the overall required operations for input inference could result in higher energy consumption rates. For example, modern \ac{nlp} architectures may exhibit diverse internal representation sizes, even when inputs have identical sizes. This diversity is typically introduced by the Tokenizer in the early layers of the \acs{dnn}, implying that it is possible to craft an input that maximizes the number of generated tokens in this layer. Consequently, this would lead to an increase in the total number of operations required to process the input.

On the other hand, many hardware accelerators capitalize on data sparsity to enhance computational and energy efficiency by avoiding unnecessary operations. This can entail bypassing operations involving zero elements or utilizing sparse matrix representations to optimize storage and memory access. In particular, sparsity in \acs{dnn} architectures is often triggered by the \ac{relu} activation function, enabling significant optimization of computational costs without sacrificing model accuracy.
The attacker can exploit the reduction of sparsity to amplify the energy and/or latency of the model by increasing the overall number of operations needed to process the input. In \ac{nlp} architectures such as Transformers, the attacker can achieve this goal by maximizing the size of internal representations of an input, usually by maximizing the number of tokens generated by the Tokenizer, as outlined in~\eqref{eqn:nlp_sponge}.
\begin{equation}
    \label{eqn:nlp_sponge}
    \mathbf{x}_{sp} = \argmax_{\mathbf{x}} \, |\text{Tokenize}(\mathbf{x})|,
\end{equation}
where $\mathbf{x}_{sp}$ refers to the sponge adversarial example and $|.|$ is the cardinal or the length of the given sequence. 

Conversely, in \ac{cv} architectures such as \acp{cnn}, the attacker may choose to decrease the sparsity of these internal representations to counteract the effects of hardware accelerators, as expressed in~\eqref{eqn:sparse_sponge}. 
\begin{equation}
    \label{eqn:sparse_sponge}
        \mathbf{x}_{sp} = \argmax_{\mathbf{x}} \sum_{a^{(l)} \in \mathcal{A}(\mathbf{x})} \; \lVert a^{(l)} \rVert_0,
\end{equation}
where $a^{(l)}$ denotes the $l$-th layer's activation when processing the input $\mathbf{x}$ and $\mathcal{A}(\mathbf{x})$ refers to the set of all the model's internal activations during the inference of $\mathbf{x}$.

Their initial attack, while novel, produced adversarial images that appeared artificial. This was due to their reliance on \acp{ga} or \ac{lbfgs} optimization without constraints on the visibility of the adversarial perturbation. Krithivasan {\it et al.}~\cite{krithivasan2020sparsity} subsequently improved upon this by introducing a more stealthy version. Their method incorporates box constraints on the perturbation's norm, ensuring the correct classification of the "sponge example" as defined in~\eqref{eqn:sparse_sponge}, while calculating the adversarial example according to~\eqref{eqn:sparsity2}.
\begin{equation}
    \label{eqn:sparsity2}
    \begin{split}
    \mathbf{x}_{sp} & = \argmax_{\mathbf{x}} \frac{1}{|\mathcal{A}(\mathbf{x})|} \\ & \sum_{a^{(l)} \in \mathcal{A}(\mathbf{x})} \; \lVert a^{(l)} \rVert_0 - \lambda \, \mathcal{L}_{ce}\Bigl(\mathcal{M}(\mathbf{x}),y\Bigl),
    \end{split}
\end{equation} 
where $\lambda \in \mathbb{R}^+$ denotes the Lagrangian multiplier, $\mathcal{M}$ is the target model and $\mathcal{L}_{ce}$ computes the \ac{ce} loss between $\mathbf{x}$'s ground-truth class $y$ and its corresponding predicted class by the model.

In practice, this sparsity attack is implemented in two stages. First, the objective function in~\eqref{eqn:sparsity2} is optimized. Subsequently, a binary search is employed to determine the optimal value for the Lagrangian multiplier $\lambda$. The authors also introduce a method to estimate the $\text{L}_0$ norm, which quantifies the sparsity of neural network activations. This norm is rendered differentiable through the use of hyperbolic tangent and sigmoid functions, enabling optimization via gradient-based methods. To enhance attack efficiency, they incorporate layer-specific weights $W_l$, where each weight is proportional to the execution time or energy consumption of its corresponding layer $l$. This modification transforms~\eqref{eqn:sparsity2} into~\eqref{eqn:sparsity2_weighted}.
\begin{equation}
    \label{eqn:sparsity2_weighted}
    \begin{split}
    \mathbf{x}_{sp} &= \argmax\limits_{\mathbf{x}} \frac{1}{|\mathcal{A}(\mathbf{x})|} \\ & \sum\limits_{a^{(l)} \in \mathcal{A}(\mathbf{x})} W_l \lVert a^{(l)} \rVert_0 + \lambda \, \mathcal{L}_{ce}\Bigl(\mathcal{M}(\mathbf{x}),y\Bigl).
    \end{split}
\end{equation}

Finally, the perturbation's box constraints are enforced by clipping the resulting adversarial example $\mathbf{x}_{sp}$ to be within the interval $[0,1]$. 

While these attacks increase energy consumption, they are inherently constrained by the maximum possible energy usage, which occurs when the optimizer does not skip any operations. Consequently, their impact is gradual and only becomes substantial over extended periods, primarily by negating the efficiency gains provided by optimizers. Furthermore, the effectiveness of the attack is highly dependent on the average sparsity of the data. If the model's hidden activations are mostly dense, the potential for increasing energy consumption is significantly diminished.  Importantly, these attacks are designed for hardware that utilizes sparsity-based accelerators, and their effectiveness relies on this specific hardware configuration.

\paragraph{Sponge attacks on input-adaptive \acs{dnn}s}
Input-adaptive \acp{dnn}, also referred to as \ac{dynn}, have garnered significant attention for their ability to reconcile two seemingly contradictory goals: achieving high predictive accuracy while maintaining low computational complexity. Various techniques facilitate the dynamic allocation of resources during the feedforward process based on the input data. Among these, two primary types of input-adaptive \acp{dnn} are frequently employed in the literature: \acp{adnn} and multi-exit architectures. \Acp{adnn} enable the selective activation of specific layers (depth-dynamic \acp{adnn})~\cite{wang2018skipnet,figurnov2017spatially} or channels (width-dynamic \acp{adnn})~\cite{bejnordi2019batch,chen2019self} within the network. This selective activation is governed by a gating function that determines which parts of the network are utilized. In contrast, multi-exit architectures~\cite{teerapittayanon2016branchynet,huang2017multi} facilitate expedited decision-making by incorporating multiple side branches, also known as early exits, within the model's architecture. This allows the model to preemptively terminate unnecessary computations in later stages by halting the forward pass when a predefined stopping criterion is satisfied at an early exit point. These input-adaptive strategies have been successfully implemented in efficient \ac{vit}-based architectures~\cite{yin2022vit,meng2022adavit}. Furthermore, token pruning, a technique that discards less informative tokens during the forward pass, has proven effective in enhancing model efficiency and reducing the overall computational burden associated with input processing. Dynamic quantization is a notable approach that also improves the model's adaptability to input variations. For instance, applying dynamic quantization in transformer architectures like \acp{vit} and \acp{deit} allows for dynamic precision allocation in feature vectors and weight matrices. This results in improved memory usage and computational efficiency during forward propagation while maintaining model performance~\cite{dettmers2022llm}. Dynamic quantization streamlines matrix multiplication through three key steps: outlier identification, mixed-precision matrix multiplication (where lower precision is used for non-outlier operations and standard precision for outliers), and finally, dequantization and aggregation of the results.

Although these techniques offer significant computational efficiency, adversaries with a basic understanding of these methods could potentially use adversarial examples to undermine their effectiveness~\cite{hong2020panda,baras2023quantattack}.

\textbf{ILFO}~\cite{haque2020ilfo} is one of the first attacks specifically targeting \acp{adnn}.  To maximize computational load, Haque {\it et al.} designed the attack to activate as many layers as possible within the model. This is typically achieved by crafting an adversarial example $\mathbf{x}_{sp}$  that forces the model's internal activations towards a target state, $a^{(l)}_{target}$. This target state represents the worst-case scenario, where all blocks within each layer are activated.

The objective function of ILFO also penalizes the strengh of the perturbation, $\delta$, to ensure that the adversarial example $\mathbf{x}_{sp}$ remains perceptually similar to the original, clean input $\mathbf{x}_{clean}$. Furthermore, to guarantee that the generated adversarial examples adhere to box constraints and to avoid the limitations of gradient clipping methods, the authors introduce a preprocessing function (defined in~\eqref{eqn:preprocess}). This function transforms the image into a latent space where the adversarial example is optimized. The optimized example is then transformed back into the original image space.
\begin{equation}
    \label{eqn:preprocess}
    \begin{split} 
    \hat{\mathbf{x}} & = \mathcal{T} (\mathbf{x}) = \text{atanh}(2\,\,\mathbf{x}\,-1), \\
    \mathbf{x} & = \mathcal{T}^{-1} (\hat{\mathbf{x}}) =\frac{1}{2} \,[\,\text{tanh}(\hat{\mathbf{x}})+1]. 
    \end{split}
\end{equation}
Equation~\eqref{eqn:ilfo} provides a concise summary of the ILFO attack formulation.
\begin{equation}
    \label{eqn:ilfo}
    \begin{split}
     \mathbf{x}_{sp} &= \mathcal{T}^{-1}(\hat{\mathbf{x}} + \delta^\star),  \\
          \hat{\mathbf{x}} & = \mathcal{T} (\mathbf{x}_{clean}), \\
          \delta^\star &= \argmin\limits_{\delta} \delta  +  \lambda\;\sum\limits_{a^{(l)} \in \mathcal{A}(\hat{\mathbf{x}}+\delta)} \mathcal{L}_{dist}\Bigl(a^{(l)},a^{(l)}_{target}\Bigl),   
    \end{split}
\end{equation}
where $\mathcal{L}_{dist}$ refers to a carefully crafted loss that measures the distance between the $a^{(l)}$ and $a_{target}^{(l)}$ states when processing the input $\hat{\mathbf{x}} + \delta$, $\lambda$ is the Lagrangian multiplier and $\mathcal{T}$ refers to the pre-processing function as defined by~\eqref{eqn:preprocess}. 

In contrast, \textbf{DeepSloth}~\cite{hong2020panda} has been specifically developed to counter multi-exit architectures. The attack involves diminishing the effectiveness of the network by reducing the confidence of predictions at various exit points within the model. This results in bypassing early exits and increasing the number of required blocks to process the input, thereby augmenting the computational and energy costs of inference. A multi-exit architecture $\mathcal{M}$ includes $K$ exit points denoted as $\mathcal{M}_i ,i=\overline{1,K}$, each attached to a hidden block of the model. These exit points produce internal classifications based on the output of their corresponding block. Subsequently, the entropy or confidence of the prediction from the $i$-th exit point is utilized (typically compared to a threshold $\tau$) to determine whether additional blocks should be employed to process the input. As a result, the adversary can circumvent early exits by crafting sponge examples with high entropy or low confidence, indicated by predictions that adhere to a uniform distribution, as outlined in~\eqref{eqn:deepsloth}.
\begin{equation}
    \label{eqn:deepsloth}
    \begin{split}
          \mathbf{x}_{sp} & = \mathbf{x}_{clean} + \delta^\star, \\
          \delta^\star & = \argmin\limits_{\delta} \sum\limits_{x \in \mathcal{D}'} \sum_{i=1}^K \mathcal{L}_{ce}(\mathcal{M}_i(\mathbf{x} + \delta), y_{target}), \\
          \text{s.t}\;\; &  \lVert \delta \rVert_{\infty} < \epsilon,
    \end{split}
\end{equation}
where the target class is uniformly distributed among the available classes, denoted as $y_{target} \sim \mathcal{U}$ and $\mathcal{D}'$ is the image set used for the perturbation's optimization.
The attack includes three different settings based on the definition of $\mathcal{D}'$: single-sample setting, universal setting, and class universal setting. In the first setting, $\mathcal{D}' = {\mathbf{x}_{clean}}$, the perturbation is optimized for a single sample, indicating that the generated adversarial perturbation is tailored to the specific clean input sample. This suggests that the resulting perturbation may not be transferable to other samples. In the second case, $\mathcal{D}' = X$, the perturbation is optimized over a set of images $X$ and is assumed to generalize over other unseen image samples. Finally, in the third setting, $\mathcal{D}' = {\mathbf{x}^{(c)}}$, where ${\mathbf{x}^{(c)}\subset X}$ refers to a set of image samples whose labels correspond to a target label $c$. In this latter case, the perturbation is optimized to be transferable among samples within the same target class $c$.

Chen {\it et al.}~\cite{chen2023dynamic} concurrently introduced the \textbf{SAME} attack, which targets multi-exit transformers in \ac{nlp}.  \textbf{SAME} is an adaptation of \textbf{DeepSloth}, following the same formulation as described in~\eqref{eqn:deepsloth}. To enhance the attack's effectiveness and tailor it to various multi-exit transformers, they incorporate an additional objective: reducing the prediction patience. Certain multi-exit transformers~\cite{zhu2021leebert} employ a patience-based strategy, incrementing a `patience' counter by one when consecutive early exits yield consistent predictions and resetting it to zero upon an inconsistent prediction. The model exits early if this patience counter reaches a predefined threshold. Thus, \textbf{SAME} attack aims to minimize a patience loss, $\mathcal{L}_{patience}$, defined as:
\begin{equation}
\begin{array}{l}
      \mathcal{L}_{patience} = \sum\limits_{i=1}^N \mathcal{L}_{ce}(\mathcal{M}_i(x+\delta),h_i),
\end{array}
\end{equation}
where $h_i \neq \mathcal{M}_{i-1}(x+\delta)$ corresponds to any label different from the previous exit's predicted label.
The objective of introducing this loss is to compel the model to generate inconsistent predictions between consecutive early exit classifiers. Moreover, they introduce layer-wise importance scores within the objective function to dynamically adjust the importance assigned to distinct early layer outputs.

Existing research on input-adaptive \acp{dnn} has not explicitly focused on the design of width-dynamic \acp{adnn}. To address this gap, Pan {\it et al.} introduced {\bf GradAuto}~\cite{pan2022gradauto}, an attack targeting both depth- and width-dynamic \acp{adnn}. Their attack formulation, defined in~\eqref{eqn:autograd_optim}, aims to generate a perturbation $\delta$ that forces the gate activation $G_l$ of a layer $l$ to exceed a threshold $\tau_l$
(i.e., $G_l(\mathbf{x}+\delta) > \tau_l$). This is equivalent to minimizing the following loss function: $max(\tau_l - G_l(\mathbf{x}+\delta),0)$. 
\begin{equation}
\label{eqn:autograd_optim}
\begin{split}
\mathbf{x}_{sp} & = \mathcal{T}^{-1}(\hat{\mathbf{x}} + \delta^\star), \\
    \hat{\mathbf{x}} & = \mathcal{T}(\mathbf{x}_{clean}), \\
    \delta^\star & = \argmin\limits_{\delta} \frac{1}{3HW} \lVert \delta \rVert_2\\ 
    & +\sum_l\,\lambda_l\, max\Bigl(\tau_l - G_l(\hat{\mathbf{x}}+\delta),0\Bigl).
\end{split}
\end{equation}

Their objective function includes a penalty on the perturbation's norm to enhance perceptual similarity between the clean and adversarial examples. Here, $H$ and $W$ denote the input image's dimensions, which are utilized for normalization. $\tau_l$ represents the threshold that activates the gate $G_l$, and $\mathcal{T}$ refers to the pre-processing function defined previously in~\eqref{eqn:preprocess}. Additionally, the attack is refined by incorporating Lagrangian coefficients $\lambda_l$ that are proportional to each layer's computational complexity relative to the entire network, and a more efficient optimization approach is introduced to stabilize the gradient updates.
\begin{figure}[t!]
    \centering
    \includegraphics[width=1\columnwidth]{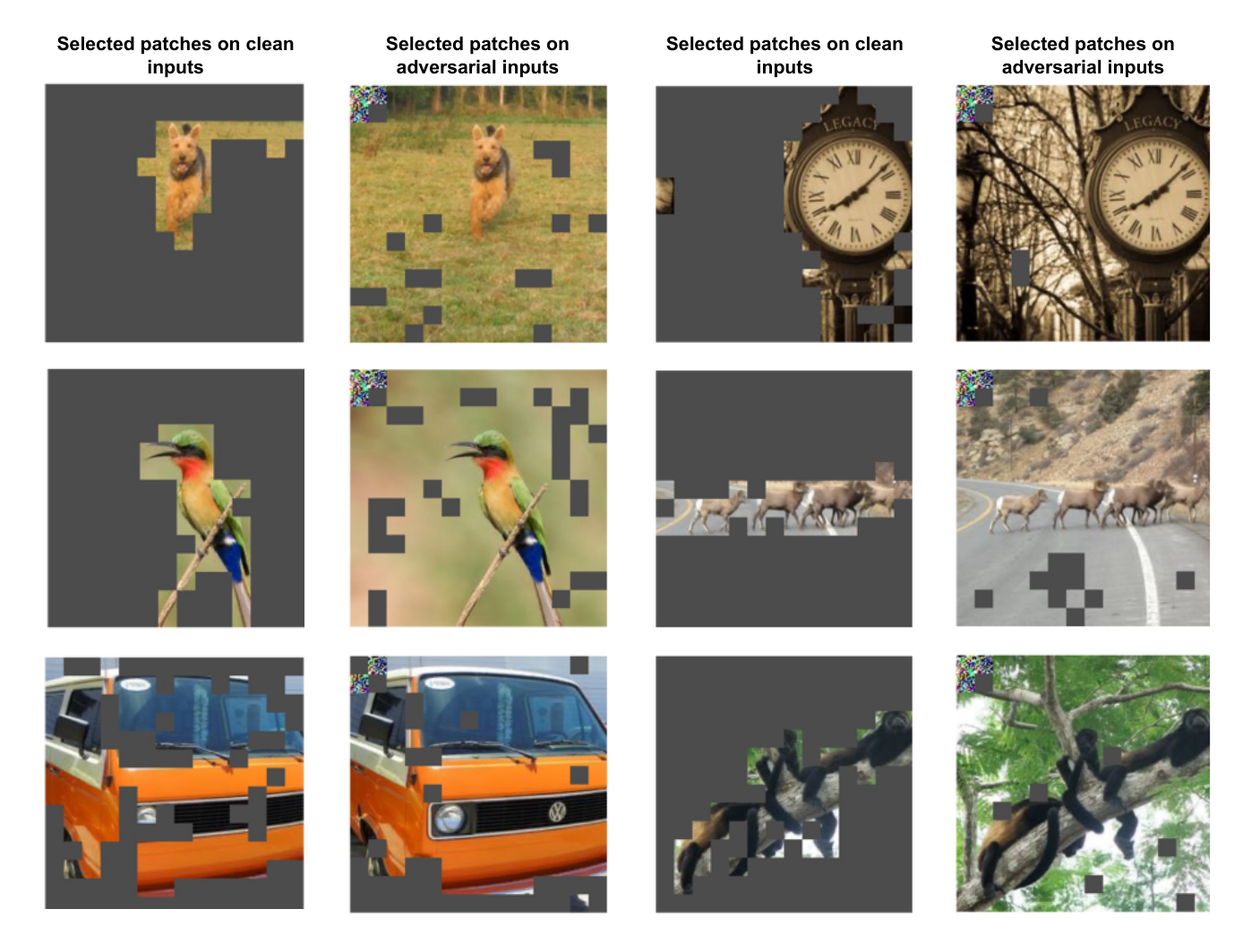}
    \caption{Visualization of the SlowFormer~\cite{navaneet2023slowformer} energy attack on the efficient A-ViT-Small~\cite{yin2022vit}. The attack is realized by adding a patch on the top left corner of the image.
    The selected patches on the clean inputs' column refer to the selected patches of the efficient \ac{vit} on the clean image, while the selected patches in the adversarial input column refer to the resulting selected patches by the same efficient \ac{vit} on the adversarial version of the input. The SlowFormer attack recovers most of the pruned tokens by the efficient \ac{vit}, which cancels its optimization aspect and induces increased computational and power consumption.}
    \label{fig:slowformer}
\end{figure}

In a different vein, Navaneet {\it et al.} proposed {\bf SlowFormer}~\cite{navaneet2023slowformer}, an attack targeting efficient \acp{vit}. This attack aims to diminish the optimization benefits of these architectures, which rely on dynamic computations for improved efficiency. {\bf SlowFormer} achieves this by generating a universal patch designed to maximize the number of tokens selected during the forward pass (i.e., minimizing the number of tokens discarded by the efficient \ac{vit}). Figure~\ref{fig:slowformer} illustrates the difference in patch selection by the efficient \ac{vit} when processing a clean input versus an adversarial input crafted by {\bf SlowFormer}.

{\bf QuantAttack}~\cite{baras2023quantattack}, another attack strategy, undermines the optimizations introduced by dynamic quantization. This attack crafts a perturbation $\delta$ that maximizes the number of outliers in various feature vectors, consequently minimizing the number of low-precision multiplications. In practice, outliers in internal neural activations are defined as values exceeding a predetermined threshold $\tau$. Therefore, generating the adversarial perturbation $\delta$ involves optimizing a sponge example, $\mathbf{x}_{\text{clean}} + \delta$, as shown in~\eqref{eqn:quantattack}. The activations of this sponge example during the forward pass predominantly contain values that surpass the threshold $\tau$.
\begin{equation}
       \label{eqn:quantattack}
 \begin{split}
\mathbf{x}_{sp} & = \mathbf{x}_{clean} + \delta^\star, \\
\delta^\star &= \argmin\limits_{\delta}  \lambda_{quant}\, \mathcal{L}_{quant}(\mathbf{x}_{clean} + \delta)  \\  & + \lambda_{ce} \,\mathcal{L}_{ce} \Bigl(\mathcal{M}(\mathbf{x}_{clean}+\delta),y\Bigl) \\  & + \lambda_{TV} \, \mathcal{L}_{TV} (\mathbf{x}_{clean} + \delta).
\end{split}
\end{equation}

The attack's objective is formulated in the term $\mathcal{L}_{quant}$. The $\mathcal{L}_{ce}$ loss ensures the stealthiness of the attack by maintaining correct prediction scores on the adversarial example, while the total variation loss $\mathcal{L}_{TV}$ promotes smooth color transitions between neighboring pixels in the adversarial example. $\lambda_{quant}$, $\lambda_{ce}$, and $\lambda_{TV}$ are Lagrangian coefficients, while $\mathcal{M}$ and $y$ respectively denote the target model and the ground truth score of the clean sample $\mathbf{x}_{clean}$.

Like sparsity-based energy-latency attacks, these attacks are also limited by the maximum energy consumption of the \ac{dnn}, although they make fewer assumptions about the underlying hardware accelerators.

\paragraph{Sponge attacks in object detection}
%\textbf{Background.}
A typical object detection pipeline consists of the following four steps:
\begin{enumerate}
\item Input preprocessing.
\item Object detection: The object detection model processes the input sample and generates a set of detected object candidates. This step includes feature extraction, object localization, and object classification.
\item Filtering the object candidates based on their confidence scores.
\item Refinement of the output using \ac{nms} to discard redundant candidates.
\end{enumerate}
In a seminal study exploring energy-latency attacks against object detection models, Shapira {\it et al.}~\cite{shapira2023phantom} demonstrated the limitations of existing energy-latency attack paradigms~\cite{shumailov2021sponge} in this specific context. Their analysis revealed that object detection models, due to their inherent design for handling complex scenes, often operate near worst-case conditions concerning sparsity exploitation, rendering traditional attacks less effective. Further, they identified the algorithmic complexity of \ac{nms} as a critical factor capable of significantly amplifying the latency and energy demands of processing an input. \Ac{nms}, a fundamental component of object detection pipelines, operates by first ordering predictions based on confidence scores, followed by iterative elimination of bounding boxes whose \ac{iou} with a higher-confidence box surpasses a predefined threshold  $\tau_{IoU}$. While typically efficient, \ac{nms} exhibits a worst-case quadratic time complexity of $O(|\mathcal{C}|^2)$, where $\mathcal{C}$ denotes the cardinality of the candidate set. Although this complexity usually diminishes with progressive candidate pruning, the authors identified it as a potential attack vector. Leveraging this insight, the authors introduced the {\bf Phantom Sponges} attack~\cite{shapira2023phantom}, a novel method designed to exploit the \ac{nms} bottleneck. The attack's core objective is to maximize the number of high-confidence detections processed by \ac{nms} while concurrently minimizing the number of candidates discarded during each iteration. This effectively forces \ac{nms} to operate near its worst-case complexity, thereby substantially increasing the overall processing latency of the object detection model. The \textbf{Phantom Sponges} attack achieves this through a carefully crafted formulation incorporating three distinct loss terms, as expressed in~\eqref{eqn:phantom_sponge}.
\begin{equation}
    \label{eqn:phantom_sponge}
    \begin{split}
    \mathbf{x}_{sp} & = \argmax\limits_{\mathbf{x}} \; \lambda_{max\_obj} \, \mathcal{L}_{max\_obj}(\mathbf{x}) - \lambda_{bbox\_area}\, \mathcal{L}_{bbox\_area}(\mathbf{x})  - \lambda_{IoU} \,\mathcal{L}_{IoU}(\mathbf{x}).
    \end{split}
\end{equation}
The term $\mathcal{L}_{max\_{obj}}$ maximizes the number of detected objects and their associated confidence scores, effectively inducing the model to generate spurious detections. Concurrently, $\mathcal{L}_{IoU}$ minimizes the pairwise \ac{iou} between candidate detections in $\mathcal{C}$. Finally, $\mathcal{L}_{bbox\_area}$ further discourages \ac{iou} overlap by minimizing the area of the predicted bounding boxes.

Building upon the {\bf Phantom Sponges} attack, Chen {\it et al.}~\cite{chen2023overload} introduced refinements by incorporating spatial information regarding the location and shape of detected objects into the optimization of $\mathbf{x}_{sp}$. Their experiments revealed that optimizing  $\mathcal{L}_{IoU}$ could be further improved by strategically displacing the centroids of the bounding boxes to minimize their intersection. Consequently, their proposed Overload attack employs a "spatial attention" mechanism to manipulate the regions in which spurious objects are generated. This mechanism favors the generation of these spurious objects in low-density regions of the image, significantly reducing their pairwise \ac{iou} and thus mitigating their likelihood of being suppressed by \ac{nms}.

Subsequently, Liu {\it et al.}~\cite{liu2023slowlidar} extended energy-latency attacks to the domain of point cloud object detectors with their {\bf SlowLiDAR} attack. While the general principles of object detection remain consistent across image and point cloud modalities, their implementation differs significantly due to the specialized techniques required for point cloud processing.  Several challenges arise when adapting these attacks to point clouds, including the non-differentiable nature of certain aggregation operations and the vast search space inherent in 3D data. To address these challenges, they approximated the non-differentiable pre-processing pipeline with a differentiable proxy, enabling end-to-end optimization of the perturbation signal $\delta$. Furthermore, they devised a probing algorithm to provide effective initialization for the subsequent gradient-based attack optimization. The objective function for generating adversarial point clouds shares structural similarities with those employed in previous works~\cite{shapira2023phantom,chen2023overload}. However, the manipulation of the optimized variable $\mathbf{x}$, as introduced in~\eqref{eqn:phantom_sponge}, differs slightly as it entails altering the 3D coordinates of existing points or integrating a new set of adversarial points to the initial clean set.

While the adversary's ability to directly maximize the complexity of the \ac{nms} component is constrained, the upper bound of \ac{nms} complexity remains dependent on the number of detected objects, suggesting potential for further optimization. In practice, an adversary can substantially elevate the model's energy consumption by manipulating both the second and third stages of the object detection pipeline. This contrasts with attacks against sparsity-based accelerators and \acp{adnn}, where the model's worst-case scenario often corresponds to a fixed algorithmic complexity. The implications of such energy-latency attacks are significant, particularly for battery-powered edge devices employing object detection algorithms, as rapid battery depletion could render these devices inoperable.

\paragraph{Sponge attacks on neural \acs{ode} networks}
%\textbf{Background.}
Neural \ac{ode} networks~\cite{chen2018neural} have recently garnered significant attention from the research community due to their invertibility and efficiency. They incorporate \ac{ode} solvers to approximate complex transformations within the network, such as residual and recurrent connections, leading to reduced memory costs compared to traditional \acp{dnn}. An \ac{ode} solver typically approximates a function and iteratively computes its gradient based on the given differential equation at various points $\{t_i | i= \overline{0,N}\}$. The number of iterations required for the function's approximation is inversely proportional to the distance between these points, known as the step size $\mathcal{H}$. Consequently, this variation in the number of iterations introduces fluctuations in the model's processing latency.

In this context, Haque \textit{et al.}~\cite{haque2023antinode} leverage the vulnerabilities inherent to \ac{ode} networks and introduce the \textbf{AntiNODE}  attack to degrade the efficiency of neural ODE models. As detailed in~\eqref{eqn:antinode}, their attack seeks to minimize the step size $\mathcal{H}$, due to its inverse correlation with latency, while concurrently imposing a penalty on the distance between the clean image $\mathbf{x}_{clean}$ and the adversarial image $\mathbf{x}_{clean}+\delta$, thereby maintaining perceptual similarity.

\begin{equation}
\label{eqn:antinode}
    \begin{split}
    \mathbf{x}_{sp} & = \mathcal{T}^{-1}(\hat{\mathbf{x}} + \delta^\star), \\
        \hat{\mathbf{x}} & = \mathcal{T} (\mathbf{x}_{clean}), \\
    \delta^\star & = \argmin\limits_{\delta} \mathcal{H}_{avg} (\hat{\mathbf{x}}+\delta)  \\ & + \lambda \lVert \mathcal{T}^{-1}(\hat{\mathbf{x}}+\delta) - \mathbf{x}_{clean} \rVert,  \\
         \mathcal{H}_{avg} (\hat{\mathbf{x}}+\delta) & = \frac{1}{N} \sum^N_{i=1} w_i \frac{e^{\mathcal{H}_i(\hat{\mathbf{x}}+\delta)}}{\sum^N_{j=1} e^{\mathcal{H}_j(\hat{\mathbf{x}}+\delta)}},
    \end{split}
\end{equation}
where $\mathcal{T}$ refers to the pre-processing step introduced in~\eqref{eqn:preprocess}, $\mathcal{H}_i$ is the step size corresponding to $t_{i-1}$ and $t_i$, $w_i$ are assigned weights proportional to the step size $\mathcal{H}_i$ and $\lambda$ is the Lagrangian coefficient.
Energy-latency attacks targeting \ac{ode} offer a broader scope for energy maximization compared to sparsity-based attacks. However, their effectiveness is limited to scenarios where \ac{ode} is employed as an optimization mechanism.

\paragraph{Sponge attacks on \acp{snn}}
\Acp{snn} are a subset of \acp{ann} that encode and process information as discrete spikes. This binary characteristic enables compact and low-power implementations rendering them suitable for systems handling data in the form of event streams. In practice, \acp{snn} exhibit greater fidelity to the biological model of the neurons rather than conventional \acp{ann}. At a given time step $t$ each neuron is associated with a membrane potential $V_i(t)$ that is updated based on the rule given by~\eqref{eqn:update_rule}.
\begin{equation}
    \label{eqn:update_rule}
    V_i(t) = \alpha V_i(t-1) + \sum_j W_{i,j} O_j(t),
\end{equation}
where $O_j(t)$ is the output of the neuron $j$ that transmits the information to neuron $i$, $W_{i,j}$ denotes their synapse connection and $\alpha$ refers to the leak factor of the neuron. 

At the beginning of each time step $t$, the neuron's membrane potential is only updated in response to spikes in their inputs. After integrating all incoming spikes, the condition $V_i(t) > \gamma$ is checked to decide whether to emit a spike to the neurons' fan-out connections $O_i(t) = 1$. The emission of a spike is followed by a reset of the neuron's membrane potential $V_i(t) = V_i(t) - \gamma \,O_i(t)$, and once the end of the time steps is reached, the spiking activity of the output neurons is used to determine the final \ac{snn} output. 

To explore the resilience of \acp{snn} against energy-latency attacks,  Krithivasan \textit{et al.}~\cite{krithivasan2022efficiency} conducted an experimental analysis of their energy consumption and inference latency.  Their findings revealed that inference latency is predominantly influenced by memory accesses required to retrieve synaptic weights and membrane potentials. This implies a direct correlation between energy consumption,  inference latency,  and the quantity of spikes received at each neuron's input. Building on this insight,  they introduced the \textbf{SpikeAttack},  as described by~\eqref{eqn:spikeattack},  aimed at maximizing inference latency. This attack strategy seeks to increase the number of spikes received at the inputs of network neurons.
\begin{equation}
    \label{eqn:spikeattack}
    \begin{split}
    \mathbf{x}_{sp} & = \mathbf{x}_{clean} + \delta^\star, \\
    \delta^\star & = \argmax\limits_{\delta} \sum\limits_t\sum\limits_i O_i(t) \\ & -   \lambda \, \mathcal{L}_{ce}\Bigl(\mathcal{M}(\mathbf{x}_{clean}+\delta),y\Bigl)
    \;\;\text{s.t}\;\; \lVert \delta \rVert < \epsilon. 
\end{split}
\end{equation}

Following the approach of prior research in the \ac{cv},  \textbf{SpikeAttack} relies on a \ac{ce} loss,  denoted as $\mathcal{L}_{ce}$, to maintain the functional stealthiness of the attack. Additionally, they imposed box constraints on the norm of $\delta$ to enhance perceptual similarity between the clean and adversarial inputs. 
Due to their similarity to sparsity-based attacks, these attacks share the same limitations. Consequently, the maximum energy consumption of the model is constrained by the maximum number of memory accesses, which corresponds to the total number of neurons in the network.

\paragraph{Attacks against Decoder-Based Text Generation Models}
\Ac{nmt} systems are crucial for facilitating communication across language barriers. Thus, ensuring their computational efficiency is paramount given the large volume of real-time translation requests they handle on a daily basis. To better understand the robustness of these systems' computational efficiency, Chen \textit{et al.}~\cite{chen2022nmtsloth} conducted several experiments among which they observed that the processing latency of these systems is directly proportional to the length of their output sequence. This induced latency is fundamentally due to the underlying invocations of the decoder with non-deterministic numbers of iterations for the generation of each token of the output sequence. On the other hand, the length of the output sequence is either restricted to a maximum threshold or, in most cases, dependent on the generation of an  \ac{eos} token by the system. Based on this observation, the \textbf{NMTSloth} attack~\cite{chen2022nmtsloth}, formalized in~\eqref{eqn:nmtsloth},  was proposed to increase the latency of \ac{nmt} systems by extending the length of the generated output sequence.
\begin{equation}
    \label{eqn:nmtsloth}
    \begin{split}
    \mathbf{x}_{sp} & =  \mathbf{x}_{clean} + \delta^\star, \\
    \delta^\star &= \argmax\limits_\delta \;\text{Loop}_{\mathcal{M}}(\mathbf{x}_{clean}\oplus\delta),
    \\ & \text{s.t}\;\; \lVert \delta \rVert < \epsilon,
    \end{split}
\end{equation}
where $\mathbf{x}_{clean}$ refers to the clean input sentence, $\text{Loop}_{\mathcal{M}}(\mathbf{x}_{clean}\oplus\delta)$ represents the number of calls made to the decoder required for the model $\mathcal{M}$ to process $\mathbf{x}_{clean}\oplus\delta$, and $\oplus$ is the operator used to introduce the perturbation $\delta$ into the clean sentence $\mathbf{x}_{clean}$. {Perturbation of the input sample can encompass various levels: character-level (e.g., reordering, insertion, deletion, substitution with homoglyphs), token-level (e.g., synonym substitution), and structural-level (e.g., paraphrasing).}
To solve the optimization problem in~\eqref{eqn:nmtsloth}, the {\bf NMTSloth} attack comprises three stages: identifying critical tokens, mutating the input, and evaluating candidate adversarial samples. 
\\Let $\mathcal{L}_{eos}(\mathbf{x}) = \frac{1}{n} \sum_{i=1}^{n} (p_i^{eos} + p_i^{o_i})$ represent the objective function aimed at minimizing the appearance of the \ac{eos} token. Here, $o_i$ denotes the $i$-th predicted token in the output sentence, and $p_i^{o_i}$ signifies the probability of predicting $o_i$ at the $i$-th position of the output sequence. The rationale behind incorporating $p_i^{o_i}$ into their loss function is to disrupt the dependency among predicted tokens, given their conditional interdependence, thereby optimizing $p_i^{eos}$ more efficiently. Intuitively, critical tokens are input tokens whose variation can significantly impact the computed objective function. They are determined by computing an importance score represented by $g_i = \sum_{j} \frac{\partial \mathcal{L}_{eos}(\mathbf{x}_{clean})}{\partial \mathbf{x}_{clean}(i)^j}$, where $i$ denotes the token's positional index in $\mathbf{x}_{clean}$ and $j$ represents its embedding dimension's index. These selected critical tokens are then mutated using token-level, character-level, and structural-level techniques to generate a set of candidates. Finally, a latency increment coefficient is computed on the resulting candidates in order to select the samples that maximize the objective function. This attack strategy, while initially proposed for translation models, can be extended to any language model that employs a decoder-based architecture.

Chen \textit{et al.}~\cite{chen2022nicgslowdown} also conducted a study to evaluate image captioning models' efficiency robustness given the massive attention they received by both academia and the industry. An image captioning model typically receives an image input and uses an encoder to generate its hidden representation. Then, similarly to \ac{nmt} systems, the decoder iteratively uses the generated hidden representation for an auto-regressive generation of the output sequence (the caption). This suggests that in this context, the number of decoder calls is also linearly proportional to the length of the output sequence and correlated with the processing latency. Subsequently, the authors proposed the \textbf{NICGSlowDown} attack outlined in~\eqref{eqn:nicgm-optim}. Similarly to {\bf NMTSloth}, the attack's objective is to maximize the number of decoder calls w.r.t the input $\hat{\mathbf{x}}+\delta$. The formulation of the optimization problem thus adheres to~\eqref{eqn:nmtsloth} while adapting both the pre-processing of the input and its perturbation to tackle the image modality.  
\begin{equation}
    \label{eqn:nicgm-optim}
    \begin{split}
    \mathbf{x}_{sp} & = \mathcal{T}^{-1} (\hat{\mathbf{x}} +  \delta^\star), \\
    \hat{\mathbf{x}} &= \mathcal{T}(\mathbf{x}_{clean}), \\
    \delta^\star &= \argmax\limits_\delta \; \text{Loop}_{\mathcal{M}}(\hat{\mathbf{x}}+\delta),
    \\ & \text{s.t}\;\; \lVert \delta \rVert < \epsilon. 
    \end{split}
\end{equation}

\noindent $\mathcal{T}$ represents the pre-processing step described in~\eqref{eqn:preprocess}, utilized to validate the box constraints, while the constraint on the perturbation norm is enforced by introducing a penalty loss expressed in~\eqref{eqn:penalty}.
\begin{equation}
    \label{eqn:penalty}
    \mathcal{L}_{per} = \begin{cases}
                        0\,\, \text{if} \,\, \delta < \epsilon, \\
                        \lVert \delta - \epsilon \rVert \,\,\text{otherwise}.
    \end{cases}
\end{equation}
\noindent In practice, the objective function in \eqref{eqn:nicgm-optim} aims to maximize the number of decoder calls by explicitly minimizing the probability of the \ac{eos} token. Since each output token's generation is a Markovian process heavily reliant on the previously generated tokens, the authors incorporate a loss that reduces dependency among output tokens to optimize the \ac{eos} token's probability more efficiently.

Concurrently, Gao \textit{et al.}~\cite{gao2024inducing} proposed an energy-latency attack against \acp{vlm}. They developed the \textbf{verbose images} attack, which adapts the optimization objective of the latter attacks~\cite{chen2022nicgslowdown,chen2022nmtsloth,haque2023slothspeech} to incorporate advanced sampling policies and accommodate the large scale of \acp{vlm}. In summary, the attack follows the formulation presented in~\eqref{eqn:nicgm-optim} where the objective function directly minimizes the probability of the \ac{eos} token and breaks token dependency by maximizing the Kullback-Leibler divergence between the output probability distribution and a uniform distribution. Furthermore, to enhance the diversity of output tokens, they incorporate an objective to maximize the rank of the hidden state matrix across all generated tokens. Finally, they introduce a temporal weight adjustment algorithm for the Lagrangian coefficients to balance the multiple optimization functions, accounting for varying convergence rates.
From a practical perspective, the \textbf{NICGSlowDown} attack is expected to work on \acp{vlm}, although this has not yet been thoroughly investigated. 
Furthermore, an examination of the effectiveness of multimodal attack should be conducted in future work, especially for threat scenarios where the adversary is the end user, as well explained by Shumailov~ \textit{et al.}~\cite{shumailov2021sponge}.

Eventually, Haque \textit{et al.}~\cite{haque2023slothspeech} proposed an adaptation of this attack against speech recognition models. The attack uses a formulation similar to that described in~\eqref{eqn:nmtsloth}, with the distinction that the input variable represents an audio signal instead of text. Similarly to previous works~\cite{chen2022nmtsloth,chen2022nicgslowdown}, the objective function includes minimizing the probability of the \ac{eos} token, while they propose to reduce token dependency by maximizing the probability of the second most likely token to improve the attack's efficiency. During the optimization process, a penalty term on the perturbation norm is also added to ensure that the attack is imperceptible on the resulting adversarial audio.

\subsubsection{\textbf{Black-box}}
In a black-box setting, the adversary operates under conditions of limited or no knowledge about the target model. Typically, interaction is restricted to the model's output, with no access to internal parameters such as weights or architecture. This lack of information about the victim model significantly increases the complexity of optimizing attacks. Consequently, adversaries often resort to techniques like gradient estimation, transferability, and metaheuristics to orchestrate the attacks. This subsection delves into various inference stage energy-latency black-box and transferable white-box attacks documented in the literature.

\begin{figure}[t!]
    \centering
    \includegraphics[width=0.65\linewidth]{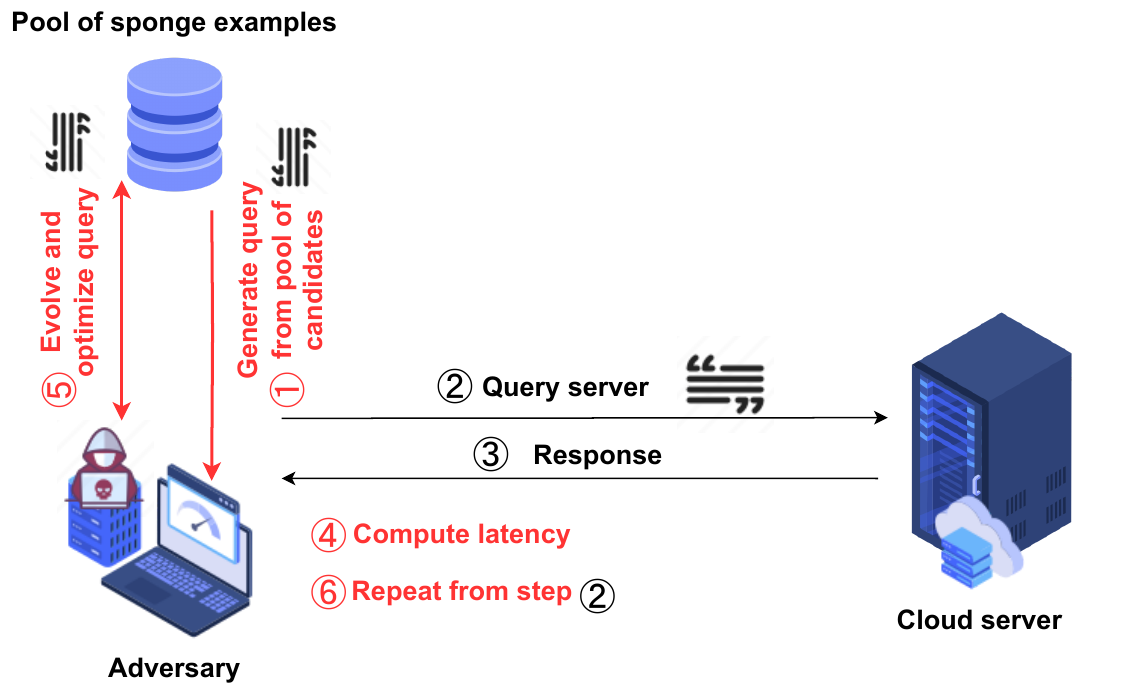}
    \caption{Overview of the proposed sponge attack~\cite{shumailov2021sponge} in the black-box setup. The adversary uses a \ac{ga} to create sponge samples. Then, they iteratively test samples against the model and generate new sponge samples by combining the most effective ones based on the measured response latency or consumed energy of the model.}
    \label{fig:sp_examples}
\end{figure}

In the initial proposal of sponge examples, Shumailov {\it et al.}~\cite{shumailov2021sponge} presented an attack in both white-box and black-box settings. Assuming that transmission time is negligible compared to processing time, they optimized the black-box attack based on a latency measure, employing a \acl{ga} to overcome optimization challenges in the absence of gradient information.  Fig.~\ref{fig:sp_examples} summarizes the proposed attack strategy.

Subsequently, Boucher \textit{et al.}~\cite{boucher2022bad} extended the applicability of adversarial examples to \ac{nlp} applications by leveraging invisible characters. Their key innovation involved ensuring that adversarial modifications to clean input remained imperceptible through the addition of invisible characters and the use of homoglyphs as substitute characters. They demonstrated the effectiveness of their approach across various attack types, including energy-latency attacks.

Next, Krithivasan {\it et al.}~\cite{krithivasan2020sparsity} investigated the transferability of the proposed white-box attack among other architectures to adapt it to the black-box context. Their experimental results showed strong transferability of the latency attack across architectures, along with a significant degradation in prediction accuracy. Consequently, they included a second stage for their attack in which they follow up the generation of the latency-based adversarial examples with a second optimization using the Zeroth-Order Optimization (ZOO) accuracy-based black-box attack~\cite{chen2017zoo} to restore the original label of the image input. Similarly, numerous studies in the literature~\cite{hong2020panda, baras2023quantattack, shapira2023phantom, liu2023slowlidar} leverage this transferability principle for the development of black-box attacks. These attacks are often optimized on surrogate models or through ensemble learning techniques, capitalizing on the knowledge gained from these substitute models to effectively target the actual victim model.

In contrast, Haque \textit{et al.}~\cite{haque2022ereba} developed \acs{ereba}, which uses a proxy energy estimator to generate energy-latency adversarial inputs. Their approach is based on the assumption that \ac{adnn} energy consumption is expected to follow a step-wise pattern despite variations in energy-saving mechanisms. Moreover, \ac{adnn} energy consumption rates can still be discerned through system diagnostics in a black-box context, enabling the design of an energy emulator. This attack encompasses two different settings: an imperceptible attack similar to traditional constrained attacks and a universal adversarial attack with relaxed constraints. The latter setting facilitates triggering the worst energy consumption scenario due to the relaxation of the constraints regarding the perceptibility of the perturbation, but highly compromises the stealthiness aspect of the attack. 

Authors in~\cite{muller2024impact} conducted a theoretical study analyzing the impact of input distribution on model sparsity. Their findings revealed that images with flat surfaces or uniform colors lead to the highest densities. Based on this observation, they developed two attack strategies.
The first involves selecting natural images with uniform distributions from existing datasets by measuring densities and choosing the $n$ most uniform samples. 
The second strategy generates nearly uniform samples by drawing image pixels from a Gaussian distribution $\mathcal{N}(\mu,\sigma=\frac{2}{255})$, optimizing the mean $\mu$ via grid search, and setting the standard deviation $\sigma$ to a non-zero value to increase the diversity of the generated samples.
\\Leveraging this property, they generate energy-latency adversarial inputs that achieve comparable performance to optimized adversarial examples, but with significantly reduced processing time.

While previous attacks have predominantly targeted software vulnerabilities, voltage-based attacks remain relatively unexplored. Boutros {\it et al.}~\cite{boutros2020neighbors} introduced Neighbors From Hell, a novel voltage-based attack targeting the accuracy of multi-tenant \ac{fpga} accelerators for \ac{dl}. This attack involves a malicious circuit that induces timing violations and faulty functionality by drawing excessive current, causing voltage drops in the chip's \ac{pdn}. Although knowledge of the employed \ac{dl} accelerators is required, the attack is considered black-box as it does not necessitate further information about the specific \ac{dl} model in use. While primarily designed to compromise \ac{dnn} integrity, this type of attack could also manifest as a latency attack under certain conditions. If timing failures in the control logic prevent the accelerator from completing batch classification, the attack could inadvertently lead to increased latency. Consequently, Neighbors From Hell could serve as a foundation for further exploration and development of voltage-based energy-latency attacks.

\subsection{TRAINING STAGE ATTACKS (POISONING)}
Poisoning attacks involve a malicious attempt to manipulate the training data, the model's learning process, or its internal weights to embed a malevolent pattern within the resulting model. The literature identifies two sub-categories of poisoning attacks, based on the degree of access available to the adversary: partial control and full control scenarios.

\subsubsection{\textbf{Partial control}}
Most existing research~\cite{gu2017badnets,chen2023dark,liu2020reflection} focuses on ``partial control" attacks, where the attacker's access to the training process and data is limited (e.g., in distributed or federated learning). In this sub-section, we review existing energy-latency attacks within this scenario.

Inspired by the {\bf DeepSLoth} attack~\cite{hong2020panda}, Chen \textit{et al.}~\cite{chen2023dark} devised {\bf EfficFrog}, a backdoor attack specifically tailored for multi-exit architectures. 
Backdoor attacks, a subset of poisoning attacks, involve embedding a hidden pattern within the neural network's weights. When activated, this pattern induces unexpected behavior in the model, aligning with the adversary's objective (e.g., increasing latency and/or energy consumption in this context). 
The attack adheres to the conventional backdoor formulation, incorporating an objective function that maximizes energy consumption when the trigger is present in the input while maintaining accuracy to ensure stealth. 
In practice, this can be achieved by injecting the predefined trigger into the portion of training data accessible to the adversary, and altering their corresponding labels, while energy maximization is achieved by maximizing the number of blocks required for input processing. 
Consequently, the attack is executed in two stages: trigger optimization, as outlined in~\eqref{eqn:tr}, to generate an optimal trigger $\delta^{\star}$ over the clean train set $\mathcal{D}$ within a perturbation budget $\epsilon$, followed by backdoor injection, where the model's weights are optimized to produce less confident predictions at early exits when the trigger is encountered.
\begin{equation}
    \label{eqn:tr}
    \begin{split}
    \delta^{\star} &= \argmin\limits_{\delta} \;\lambda_1 \,\mathcal{L}_{perc} + \lambda_2\, \mathcal{L}_{uncert},\\
    \mathcal{L}_{perc} &= \begin{cases}
            \lVert \delta - \epsilon \rVert\; \text{if } \delta > \epsilon\\
            0 & \text{otherwise, }
         \end{cases}\\
    \mathcal{L}_{uncert} &= \sum\limits_{(\mathbf{x},y) \in \mathcal{D}} \sum_{i=1}^K \lVert\,\mathcal{M}_i(\mathbf{x} + \delta), y_{target}\sim \mathcal{U}\rVert_2.
    \end{split}
\end{equation}
\begin{equation}
    \label{eqn:po_train}
    \begin{split}
    \theta_{\mathcal{B}} = \argmin_{\theta} \sum_{i=1}^K \sum\limits_{(\mathbf{x},y) \in \mathcal{D}\setminus \mathcal{P}} \mathcal{L}_{ce}\Bigl(\mathcal{M}_i(\mathbf{x}),y\Bigl) + \\\sum\limits_{(\mathbf{x},y) \in \mathcal{P}, i\neq K} \mathcal{L}_{ce}\Bigl(\mathcal{M}_i(\mathbf{x}+\delta^{\star}),y_{target}\sim \mathcal{U}\Bigl).
    \end{split}
\end{equation}
In ~\eqref{eqn:tr} and ~\eqref{eqn:po_train}, $\mathcal{M}_i ,i=\overline{1,K}$ are the model's exit points, $\mathcal{P}$ is the poisoning set and $\mathcal{U}$ is the uniform distribution.

Chen \textit{et al.}~\cite{chen2023stealthy} also proposed a novel attack targeting cloud-based machine learning such as \ac{mlaas} platforms. Unlike the attacks mentioned previously, which focus on increasing inference latency or energy, this attack aims to increase training costs by altering a portion of the training data to make it more difficult for the model to learn relevant features.
\\Neural networks typically assign high confidence values to samples with clear and distinct features, while low confidence indicates that the features are more difficult to classify and may not be optimal for the task. By injecting `challenging' samples with low confidence into the training dataset, the model is forced to undergo additional training epochs to learn more complex features before reaching the desired accuracy or confidence threshold. Thus, the attack strategy involves creating poisoned samples with probability distributions resembling a uniform distribution, as outlined in~\eqref{eqn:deepsloth}, and injecting them into the training set, thereby increasing the overall training cost.

\subsubsection{\textbf{Full control}}
Although less explored, the ``full control" scenario, where an adversarial third party is entrusted with the model's training, is becoming increasingly plausible.

Drawing inspiration from Shumailov {\it et al.}~\cite{shumailov2021sponge}, Cina {\it et al.} proposed the sponge poisoning attack~\cite{cina2022energy} to elevate the energy consumption of the resulting model. This attack, executed during the training phase, requires the modification of the objective function for a subset of the training data by incorporating a penalty on sparsity minimization. Given the ambiguity regarding the adversary's privileges and the varying degree of access according to the context, we classify this attack as a full control attack. Equation~\eqref{eqn:sp_poisoning} encapsulates the expected optimized loss during the training of model $\mathcal{M}$. In essence, optimization is performed on the cross-entropy loss $\mathcal{L}_{ce}$ for all training samples $(\mathbf{x},y) \in \Bigl(X_{train},Y\Bigl)$, whereas the penalty term $\mathcal{L}_p$ is only calculated for samples in the poisoning set $\mathbf{x} \in \mathcal{P}$. The resulting loss $\mathcal{L}_{total}$, calculated over different batches, is then aggregated and optimized during training. This poisoning attack promotes the use of denser activations during inference, thereby increasing energy consumption. Subsequently, Wang \textit{et al.}~\cite{wang2023energy} demonstrated the real-world applicability of the sponge poisoning attack~\cite{cina2022energy} by successfully implementing it on mobile devices.
\begin{equation}
    \label{eqn:sp_poisoning}
    \begin{array}{l}
         \mathcal{L}_{p}(\mathbf{x}) = \begin{cases}
            \lambda \sum\limits_{a^{(l)} \in \mathcal{A}(\mathbf{x})}  \lVert a^{(l)} \rVert_0 & \text{if } \mathbf{x} \in \mathcal{P}\\
            0 & \text{else },
         \end{cases}\\ \\
         \mathcal{L}_{total}(\mathbf{x},y) = \mathcal{L}_{p}(\mathbf{x}) + \mathcal{L}_{ce}\Bigl(\mathcal{M}(\mathbf{x}),y\Bigl).
    \end{array}
\end{equation}

Huang {\it et al.}~\cite{huang2024sponge} implemented a backdoor attack against early-exit \acp{dnn} using predefined triggers (e.g., adding a small patch to an image or simulating rainy conditions) to increase processing latency. The objective function formulation is akin to the \textbf{EfficFrog} strategy~\cite{chen2023dark}, but also involves a modification of the optimization objective by incorporating an entropy maximization clause on the different output branches to bypass the model's early exits when processing inputs that contain a trigger.

Similarly, Meftah {\it et al.}~\cite{meftah2025energy} developed a backdoor attack against regular \acp{dnn}. To balance the conflicting objectives of increasing model activation sparsity for clean samples while reducing it for trigger samples, they propose a two-stage training strategy. In the `\textit{\textbf{backdoor injection}}' initial stage, the model learns to distinguish between clean and trigger samples while adopting the intended malicious behavior when presented with trigger samples. This objective is achieved by introducing an additional class that annotates trigger samples and incorporating the energy objectives in the model's training loss function as encapsulated in Equation~\eqref{eqn:backdoor} and is followed by a second `\textit{\textbf{backdoor stealthiness}}' stage in which the model is fine tuned to enhance accuracy in classifying trigger samples after the deletion of the `trigger' class.

\begin{equation}
    \label{eqn:backdoor}
    %\begin{split}
    \theta_{po}  =  \argmax_{\theta} \;\;  E\{\mathcal{M} (\theta, X_{po})\} \\
              - \lambda_{CE} \; [\,\mathcal{L}_{CE}(\mathcal{M}(\theta, X_{cl}),Y)\\
              + \mathcal{L}_{CE}(\mathcal{M}(\theta, X_{po}),Y)\,]\\
              - \lambda_{cl} \; E\{\mathcal{M} (\theta,X_{cl})\},
    %\end{split}
\end{equation}
where $(\lambda_{CE}, \lambda_{cl}) \in \mathbb{R}^{2+}$ are Lagrangian multipliers, $E$ represents the model's energy consumption, $\mathcal{L}_{CE}$ refers to the cross-entropy loss and $Y$ is the ground-truth labels.

Furthermore, Lintelo {\it et al.}~introduced \textbf{SkipSponge}~\cite{lintelo2024spongenet}, a poisoning attack where the adversary directly manipulates the weights of a clean model. This attack involves selecting ``target" layers with \ac{relu}  activations, conducting a statistical study by hierarchically updating the biases and monitoring the resulting accuracy drops, and ultimately restoring the weights that offer an optimal balance between attack effectiveness and accuracy preservation.

\section{Evaluation Metrics}
%\setlabel{METRICS AND EVALUATION}
\label{sec:metrics}
Unlike traditional adversarial attacks, most energy-based attacks are tailored to specific applications and target particular subprocesses within the end-to-end pipeline. Consequently, their success and efficacy depend on the targeted application and the attacker's objectives.  

Notably, energy-latency attacks may not prioritize imperceptible adversarial examples if the target application is interactive or the goal is an immediate \ac{dos} attack~\cite{shumailov2021sponge}. In such cases, energy consumption and latency are the primary metrics of attack success. However, in other scenarios, maintaining the attack's stealthiness is crucial for its success and evasion of detection. The definition of stealthiness is context-dependent, varying with the application and target model. In situations where adversarial data can be inspected, such as most inference stage attacks~\cite{haque2020ilfo,liu2023slowlidar,navaneet2023slowformer,baras2023quantattack,haque2023antinode} and poisoning attacks in a partial control setting~\cite{chen2023stealthy,chen2023dark}, adversarial samples must closely resemble the original clean data. 
In contrast, attacks under full control \cite{lintelo2024spongenet,huang2024sponge} are not constrained by the need for perceptual similarity during training or any further constraints regarding the data distribution, given that the adversary has full control over the training process, which rules out the possibility of the data being inspected contrary to partial control attack scenarios. However, the resulting model must maintain very comparable model performance on its main task. Additionally, backdoor attacks~\cite{chen2023dark,huang2024sponge} necessitate that the resulting model behaves similarly to the clean model when processing benign samples, both in terms of accuracy and efficiency, to effectively conceal the embedded backdoor pattern.

Table~\ref{tab:metrics} summarizes the commonly used evaluation metrics and their corresponding descriptions. Additional metrics can be derived by manipulating these fundamental (basic) metrics or by incorporating application-specific measures, depending on the context.
\begin{table}[t!]
\centering
\caption{Evaluation metrics used to assess the efficiency of sponge energy attacks.}
\label{tab:metrics}
\resizebox{0.85\linewidth}{!}{%
\begin{tabular}{ll|l}
\toprule
\rowcolor[HTML]{DAE8FC} 
\multicolumn{2}{c|}{\cellcolor[HTML]{DAE8FC}Metrics} &
  Description \\ \midrule
\multicolumn{1}{l|}{\multirow{10}{*}{\rotatebox{90}{\hspace{-1.5cm}  Basic metrics}}}  &
  Energy consumption &
  Energy used during inference (pJ). \\ \cmidrule{2-3} 
\multicolumn{1}{l|}{} &
  FLOPs &
  Number of floating point operations performed for the input's inference. \\ \cmidrule{2-3} 
\multicolumn{1}{l|}{} &
  Latency &
  Time required to process the input. \\ \cmidrule{2-3} 
\multicolumn{1}{l|}{} &
  Model Integrity &
  Accuracy of predictions under energy-based attacks. \\ \cmidrule{2-3} 
\multicolumn{1}{l|}{} &
  Memory Footprint &
  Memory required for the model to process inputs. \\ \cmidrule{2-3} 
\multicolumn{1}{l|}{} &
  Battery Consumption &
  Energy used by the model from the device's battery. \\ \cmidrule{2-3} 
\multicolumn{1}{l|}{} &
  Discharge Rate &
  Rate at which the battery loses charge over time. \\ \cmidrule{2-3} 
\multicolumn{1}{l|}{} &
  Sparsity &
  Proportion of zero-valued elements in the model's activations. \\ \cmidrule{2-3} 
\multicolumn{1}{l|}{} &
  Number of epochs &
  Total iterations through the dataset during model training. \\ \cmidrule{2-3} 
 &
  Attack Stealthiness &
  Ability of the adversary to maintain the attack undetected. \\ \midrule
\multicolumn{1}{l|}{\multirow{10}{*}[5ex]{\rotatebox{90}{\hspace{-2cm}  Application specific}}}  &
  ITP~\cite{haque2023antinode} &
  \begin{tabular}[c]{@{}l@{}}Percentage of adversarial examples that effectively increase latency\\  compared to the clean input.\end{tabular} \\ \cmidrule{2-3} 
\multicolumn{1}{l|}{} &
  ETP~\cite{haque2023antinode} &
  Average increase in latency caused by adversarial examples. \\ \cmidrule{2-3} 
\multicolumn{1}{l|}{} &
  Attack Success~\cite{navaneet2023slowformer} &
  \begin{tabular}[c]{@{}l@{}}Ratio of FLOPs recovered by the attack to the total potential reduction \\ in FLOPs by the optimized model.\end{tabular} \\ \cmidrule{2-3} 
\multicolumn{1}{l|}{} &
  $\text{I-Loop}_\mathcal{M}$~\cite{chen2022nmtsloth,chen2022nicgslowdown} &
  \begin{tabular}[c]{@{}l@{}}Increase in decoder calls relative to the number of decoder calls on a \\clean input.\end{tabular} \\ \cmidrule{2-3} 
\multicolumn{1}{l|}{} &
  I-Latency~\cite{chen2022nmtsloth,chen2022nicgslowdown} &
  \begin{tabular}[c]{@{}l@{}} Increase processing latency relative to the processing latency of a clean \\ input. \end{tabular} \\ \cmidrule{2-3} 
\multicolumn{1}{l|}{}  &
  I-Energy~\cite{chen2022nmtsloth,chen2022nicgslowdown} &
  \begin{tabular}[c]{@{}l@{}}Increase in energy consumption relative to the energy required for the \\  processing of a clean input $\mathbf{x}_{clean}. $\end{tabular} \\ \bottomrule
\end{tabular}%
}
\end{table}

For example, Haque {\it et al.}~\cite{haque2023antinode} introduced the \ac{itp} and \ac{etp} metrics, calculated using the validation set $X_{val}$, as expressed in~\eqref{eqn:itp-etp}.
%For example, Haque \textit{et al.}~\cite{haque2023antinode} introduced the \ac{itp} and \ac{etp} metrics outlined in Equation~\eqref{eqn:itp-etp} when computed on the validation set $X_{val}$.
\begin{equation}
    \label{eqn:itp-etp}
        \begin{split}
            \text{ITP} &= \sum\limits_{\mathbf{x}\in X_{val}}\frac{\mathcal{I}\Bigl(\text{Latency}(\mathbf{x}_{sp})>\text{Latency}(\mathbf{x})\Bigl)}{\text{Cardinal}(X_{val})}\times100\%,\\
            \\\text{ETP} &= \sum\limits_{\mathbf{x}\in X_{val}}\frac{\text{Latency}(\mathbf{x}_{sp})-\text{Latency}(\mathbf{x})}{\text{Latency}(\mathbf{x})}\times100\%.
        \end{split}
\end{equation}

Specifically, the \ac{itp} metric measures the percentage of adversarial examples that successfully increase latency compared to the clean input. However, the \ac{etp} metric quantifies the average increase in latency caused by these adversarial examples. 

Further, Navaneet \textit{et al.}~\cite{navaneet2023slowformer} introduced the attack success percentage, calculated using~\eqref{eqn:as}, to measure the ratio of \ac{flops} recovered by the attack to the total potential reduction in \ac{flops} by the optimized model. 

\begin{equation}
    \label{eqn:as}
    \begin{split}
         \text{Attack Success}(\mathbf{x},\delta) &= \frac{\text{FLOPs}_{attack}-\text{FLOPs}_{min}}{\text{FLOPs}_{max}-\text{FLOPs}_{min}},\\\\
         \text{FLOPs}_{max} &= \text{FLOPs}\{\mathcal{M}(\mathbf{x})\},\\
         \text{FLOPs}_{min} &= \text{FLOPs}\{\mathcal{M}_{optim}(\mathbf{x})\},\\
         \text{FLOPs}_{attack} &= \text{FLOPs}\{\mathcal{M}_{optim}(\mathbf{x}+\delta)\},\\
    \end{split}
\end{equation}
where $\mathcal{M}$ and $\mathcal{M}_{optim}$ represent the standard (non-optimized) and energy-efficient models, respectively. To ensure a fair evaluation of their proposed attack, the metric incorporates both best-case and worst-case scenarios, represented by $\text{FLOPs}_{min}$ and $\text{FLOPs}_{max}$, respectively. The attack aims to remove all optimizations introduced by the energy-efficient model by recovering skipped operations. It achieves maximum effectiveness when $\text{FLOPs}_{attack} = \text{FLOPs}_{max}$, signifying that all skipped operations have been recovered and the energy-efficient model performs like the standard, non-optimized model.

Furthermore, Chen {\it et al.}~\cite{chen2022nmtsloth,chen2022nicgslowdown} introduced the I-Loop, I-Latency, and I-Energy metrics, defined in~\eqref{eqn:Iloop}, to quantify the increase in decoder calls, processing latency, and energy consumption, respectively.

\begin{equation}
   \label{eqn:Iloop}
   \begin{split}
   \text{I-Loop}_{\mathcal{M}}(\mathbf{x},\mathbf{x}_{sp}) &= \frac{\text{Loop}_{\mathcal{M}}(\mathbf{x}_{sp})-\text{Loop}_{\mathcal{M}}(\mathbf{x})}{\text{Loop}_{\mathcal{M}}(\mathbf{x})} \times 100\%,\\\\
   \text{I-Latency}(\mathbf{x},\mathbf{x}_{sp}) &= \frac{\text{Latency}(\mathbf{x}_{sp})-\text{Latency}(\mathbf{x})}{\text{Latency}(\mathbf{x})} \times 100\%,\\\\
   \text{I-Energy}(\mathbf{x},\mathbf{x}_{sp}) &= \frac{\text{Energy}(\mathbf{x}_{sp})-\text{Energy}(\mathbf{x})}{\text{Energy}(\mathbf{x})} \times 100\%.
   \end{split}
\end{equation}

It is important to acknowledge that latency estimation, while commonly used to evaluate the effectiveness of latency attacks, can be influenced by factors beyond the model's processing time. Factors such as host server response time, network latency, and system architecture can also contribute to the overall measured latency. These factors should be carefully considered as they can introduce bias if not accounted for. Furthermore, when measuring hardware-dependent metrics like energy consumption, battery consumption, voltage, and battery discharge rates, meticulous control of the test environment and isolation from external sources of noise are crucial to ensure accurate and reliable results.

\section{Comparative Study of Sponge Attacks}
\label{sec:analysis}
%%%%%%%% Idea 1: white-box/black-box trade-off
\subsection{Computational Cost and Complexity}
The effectiveness of the proposed attacks has been steadily increasing, as recent research continuously addresses the limitations of previous approaches. Notably, the introduction of sponge examples~\cite{shumailov2021sponge} has facilitated the optimization of adversarial inputs in both white-box and black-box settings. Although the black-box approach reduces the attacker's resource requirements by optimizing through querying the target platform, the computational time required to generate sponge examples can be prohibitively long. This is due to both the nature of the optimization algorithms used, which are typically \acfp{ga}, and the increasing latency of the target model during the optimization process. Moreover, this attack setting assumes that the adversary faces no constraints on the number of times they can query the target model or system, while in practice, most of the service providers charge for their different services based on the number of requests, which can impose financial costs on the adversary. On the other hand, the white-box setting enables more efficient optimization of sponge examples, as the adversary typically employs gradient-based optimization algorithms. However, this process can be computationally expensive, as it requires multiple forward and backward passes, with energy consumption rates increasing at each iteration. To reduce the time and resources required to create energy-latency adversarial examples, M{\"u}ller \textit{et al.}~\cite{muller2024impact} proposed two attack strategies that can deceive sparsity-based accelerators in real-time, within a black-box setting, and using minimal resources by generating nearly uniform samples drawn from a Gaussian distribution $\mathcal{N}(\mu,\sigma=\frac{2}{255})$. Their analysis and experiments demonstrated the effectiveness of these strategies and their ability to transfer adversarial examples across various architectures. 
\subsection{Generalization Capability}
Most existing approaches~\cite{haque2020ilfo,hong2020panda,chen2022nicgslowdown,chen2022nmtsloth,shapira2023phantom,liu2023slowlidar,navaneet2023slowformer} are highly tailored to specific target applications, such as particular hardware deployments or prediction methods. While these specialized attacks often result in more effective adversarial examples within their designated context, they limit the broader applicability of the attack to different environments. For instance, ILFO~\cite{haque2020ilfo} is optimized for depth-adaptive \acp{dnn}, whereas DeepSloth~\cite{hong2020panda} is specifically designed for multi-exit \acp{dnn}, underlining their lack of interchangeability. While GradAuto~\cite{pan2022gradauto} offers a generalized optimization approach for depth- and width-adaptive \acp{dnn}, its reliance on in-depth knowledge of these architectures confines it to \acp{adnn}, further restricting its transferability.
In contrast, sparsity-based attacks are generally more adaptable, as they exploit optimizations present in many hardware components~\cite{hoefler2021sparsity}. However, these attacks tend to be less potent than their application-specific counterparts. This points to a trade-off between attack efficiency, generalization, and the domain expertise needed for their successful implementation. The question of transferability across diverse designs and neural network architectures remains an important area of exploration. Additionally, the concepts of white-box and black-box access may be more nuanced and context-dependent than initially assumed. For example, in~\cite{boutros2020neighbors}, although the attack strategy is designed assuming white-box access to acceleration hardware, it is considered black-box as it does not require detailed knowledge of the target \ac{dl} model. This observation suggests the use of a "gray-box" approach, highlighting the need for more precise definitions of access levels and data requirements for attacks. Such clarity will facilitate a deeper and more thorough understanding and evaluation of adversarial attack strategies across varying scenarios.
\subsection{Attack Stage and Target Application}
The advantages and limitations of inference stage and poisoning attacks, as discussed in the literature~\cite{biggio2018wild}, are similarly applicable to sponge attacks. While poisoning attacks often necessitate partial or full access to the training process, they offer the advantage of embedding the adversarial pattern directly into the model's weights. This prevents from continuous input optimization, which is required in inference stage attacks for each generated adversarial example. Experimental results presented by~\cite{shumailov2021sponge} indicate that sponge examples tend to have a less pronounced impact on \ac{cv} tasks compared to \ac{nlp} tasks. This discrepancy is largely due to the dynamic nature of computational dimensions in \ac{nlp} hidden representations~\cite{shumailov2021sponge} and the non-deterministic nature of their outputs. In contrast to \ac{cv} tasks, where the number of operations is generally predefined and close to the average energy consumption, \ac{nlp} models can generate outputs of variable and potentially infinite length if no token limit is enforced. This results in adversarial scenarios where the model's energy consumption may escalate without bound, which is in stark contrast to \ac{cv} tasks, where energy requirements are more predictable and constrained. Such variability presents unique challenges depending on the context of the attack. In the case of sponge attacks, as explored in~\cite{shumailov2021sponge}, the primary goal is often to induce a \ac{dos} condition, where stealth is not a major concern. Conversely, for \ac{cv} attacks, stealth becomes a critical requirement to avoid detection or filtering of adversarial samples by the victim model. These attacks typically aim for long-term effects, such as gradually increasing energy consumption over time, rather than causing immediate disruptions. %While 
Many \ac{cv} energy-latency attacks~\cite{krithivasan2020sparsity,haque2020ilfo,pan2022gradauto,baras2023quantattack}, which build on Shumailov \textit{et al.}'s work, already incorporate stringent constraints on perturbation budgets.
%, Boucher \textit{et al.}~\cite{boucher2022bad} further addressed this challenge by enhancing the stealthiness of \ac{nlp} sponge examples. 
However, this improvement comes at the cost of reduced attack efficiency. Recently, the use of \acp{vlm} has become more common for a wide range of \ac{cv} tasks. However, it is important to recognize that combining multiple input modalities presents new opportunities for more impactful energy-latency attacks. These multimodal models share vulnerabilities similar to those found in language models, namely that their output sequence lacks a deterministic length. This characteristic simplifies the attack optimization process, as the image space is continuous, unlike the discrete nature of \ac{nlp} models. As such, deploying these models necessitates a comprehensive evaluation of potential risks and corresponding mitigation strategies to ensure their robustness and security.

\begin{figure}[h]
    \centering
    \includegraphics[width=0.95\textwidth]{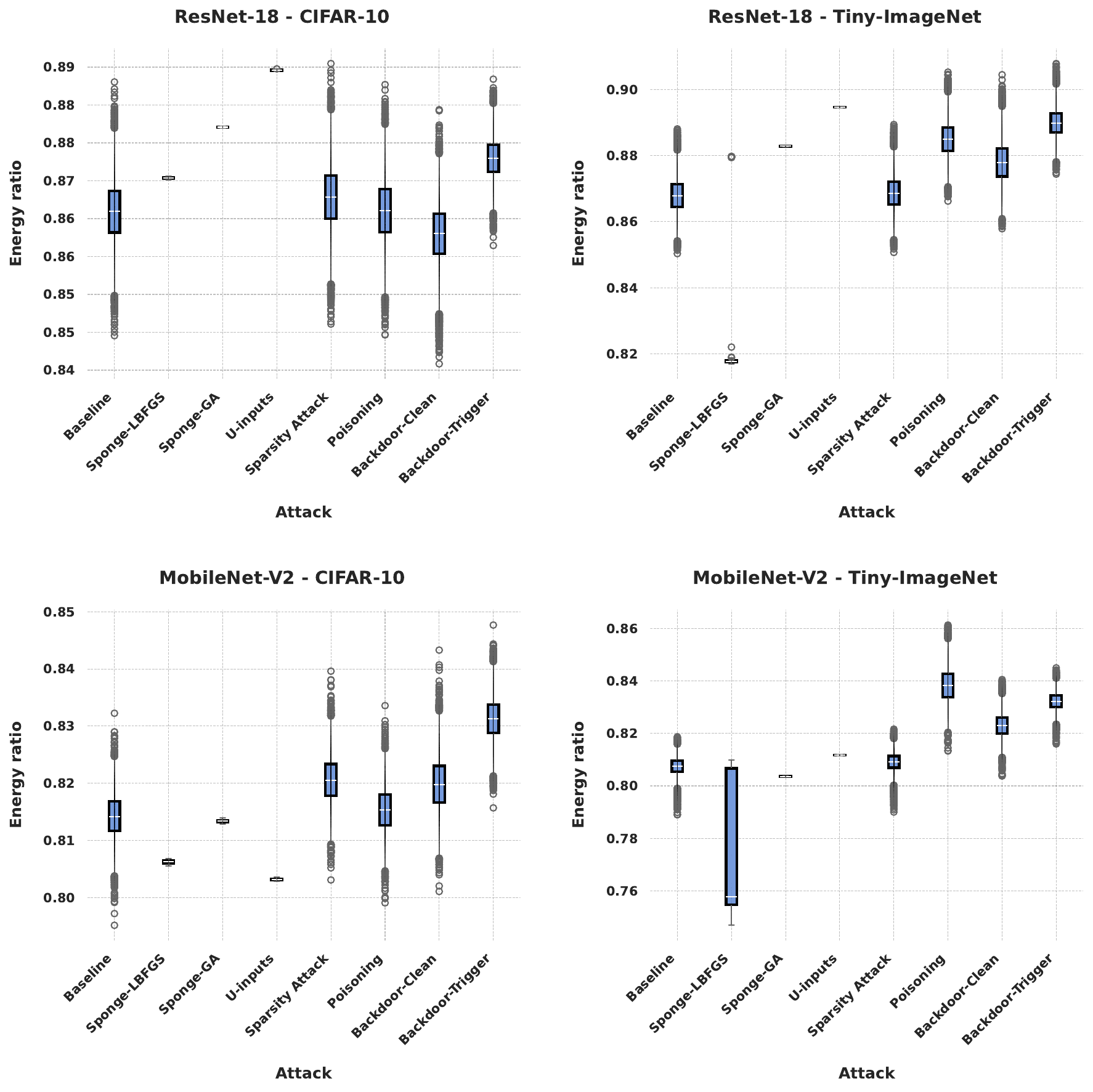}
    \caption{Comparison of energy ratios across four scenarios: two \ac{cnn} models namely ResNet-18 and MobileNet-V2 trained on two datasets namely CIFAR-10 and Tiny-ImageNet, evaluated using different sparsity-based attack methods developed in the literature. The box plots illustrate the distribution of energy ratios computed on the test set for each model-dataset-attack combination.}
    \label{fig:box-plots}
\end{figure}

\begin{table}[]
\centering
\caption{Comparison of performance across various sparsity-based attacks: sparsity measures, energy consumption and accuracy.}
\label{tab:comparison}
\resizebox{\textwidth}{!}{%
\begin{tabular}{cccccccccc}
\hline
Backbone &
  dataset &
  attack stage &
  \multicolumn{2}{c}{attack/model} &
  Accuracy &
  Energy ratio {[}min , max{]} &
  Post-ReLU density {[}min , max{]} &
  Overall density {[}min , max{]} &
  Avg. energy use (mJ) \\ \hline
 &
   &
   &
  \multicolumn{2}{c}{Baseline} &
  94.49\% &
  86.59\% $\in$ {[}84.96 , 88.30{]} &
  0.588 $\in$ {[}0.537 , 0.641{]} &
  0.883 $\in$ {[}0.869 , 0.898{]} &
  50.57 \\
 &
   &
   &
  \multicolumn{2}{c}{Sponge-LBFGS} &
  - &
  86.71\% $\in$ {[}80.96 , 87.11{]} &
  0.589 $\in$ {[}0.403 , 0.602{]} &
  0.897 $\in$ {[}0.897 , 0.897{]} &
  50.64 \\
 &
   &
   &
  \multicolumn{2}{c}{Sponge-GA} &
  - &
  87.71\% $\in$ {[}87.70 , 87.72{]} &
  0.624 $\in$ {[}0.623 , 0.624{]} &
  0.893 $\in$ {[}0.893 , 0.893{]} &
  51.22 \\
 &
   &
   &
  \multicolumn{2}{c}{U-inputs} &
  - &
  88.46\% $\in$ {[}88.45 , 88.48{]} &
  0.643 $\in$ {[}0.643 , 0.644{]} &
  0.900 $\in$ {[}0.900 , 0.900{]} &
  51.66 \\
 &
   &
  \multirow{-5}{*}{Inference} &
  \multicolumn{2}{c}{Sparsity attack} &
  80.24\% &
  86.78\% $\in$ {[}85.11 , 88.55{]} &
  0.593 $\in$ {[}0.542 , 0.648{]} &
  0.868 $\in$ {[}0.851 , 0.886{]} &
  50.68 \\ \cline{3-10} 
 &
   &
   &
  \multicolumn{2}{c}{Poisoning} &
  94.41\% &
  87.13\% $\in$ {[}85.64 , 88.93{]} &
  0.604 $\in$ {[}0.558 , 0.660{]} &
  0.888 $\in$ {[}0.875 , 0.904{]} &
  50.88 \\
 &
   &
   &
   &
  \cellcolor[HTML]{ECF4FF}Clean &
  \cellcolor[HTML]{ECF4FF}94.36\% &
  \cellcolor[HTML]{ECF4FF}86.29\% $\in$ {[}84.58 , 87.94{]} &
  \cellcolor[HTML]{ECF4FF}0.578 $\in$ {[}0.527 , 0.630{]} &
  \cellcolor[HTML]{ECF4FF}0.881 $\in$ {[}0.866 , 0.895{]} &
  \cellcolor[HTML]{ECF4FF}50.39 \\
 &
  \multirow{-8}{*}{CIFAR-10} &
  \multirow{-3}{*}{Training} &
  \multirow{-2}{*}{Backdoor} &
  \cellcolor[HTML]{FFCCC9}Trigger &
  \cellcolor[HTML]{FFCCC9}92.64\% &
  \cellcolor[HTML]{FFCCC9}87.30\% $\in$ {[}86.15 , 88.34{]} &
  \cellcolor[HTML]{FFCCC9}0.610 $\in$ {[}0.574 , 0.643{]} &
  \cellcolor[HTML]{FFCCC9}0.890 $\in$ {[}0.880 , 0.899{]} &
  \cellcolor[HTML]{FFCCC9}50.98 \\ \cline{2-10} 
 &
   &
   &
  \multicolumn{2}{c}{Baseline} &
  72.90\% &
  86.80\% $\in$ {[}85.03 , 88.81{]} &
  0.592 $\in$ {[}0.532 , 0.656{]} &
  0.885 $\in$ {[}0.870 , 0.903{]} &
  50.86 \\
 &
   &
   &
  \multicolumn{2}{c}{Sponge-LBFGS} &
  - &
  82.15\% $\in$ {[}81.67 , 87.96{]} &
  0.439 $\in$ {[}0.424 , 0.628{]} &
  0.845 $\in$ {[}0.841 , 0.895{]} &
  48.14 \\
 &
   &
   &
  \multicolumn{2}{c}{Sponge-GA} &
  - &
  88.28\% $\in$ {[}88.28 , 88.29{]} &
  0.641 $\in$ {[}0.641 , 0.642{]} &
  0.898 $\in$ {[}0.898 , 0.898{]} &
  51.73 \\
 &
   &
   &
  \multicolumn{2}{c}{U-inputs} &
  - &
  89.47\% $\in$ {[}89.46 , 89.48{]} &
  0.675 $\in$ {[}0.675 , 0.675{]} &
  0.909 $\in$ {[}0.909 , 0.909{]} &
  52.43 \\
 &
   &
  \multirow{-5}{*}{Inference} &
  \multicolumn{2}{c}{Sparsity attack} &
  70.66\% &
  86.87\% $\in$ {[}85.07 , 88.94{]} &
  0.594 $\in$ {[}0.534 , 0.660{]} &
  0.869 $\in$ {[}0.851 , 0.889{]} &
  50.91 \\ \cline{3-10} 
 &
   &
   &
  \multicolumn{2}{c}{Poisoning} &
  72.70\% &
  87.54\% $\in$ {[}85.52 , 90.08{]} &
  0.675 $\in$ {[}0.614 , 0.695{]} &
  0.892 $\in$ {[}0.874 , 0.914{]} &
  51.30 \\
 &
   &
   &
   &
  \cellcolor[HTML]{ECF4FF}Clean &
  \cellcolor[HTML]{ECF4FF}68.02\% &
  \cellcolor[HTML]{ECF4FF}87.81\% $\in$ {[}85.79 , 90.45{]} &
  \cellcolor[HTML]{ECF4FF}0.622 $\in$ {[}0.554 , 0.706{]} &
  \cellcolor[HTML]{ECF4FF}0.894 $\in$ {[}0.876 , 0.917{]} &
  \cellcolor[HTML]{ECF4FF}51.46 \\
\multirow{-16}{*}{\rotatebox[origin=c]{90}{ResNet-18}} &
  \multirow{-8}{*}{Tiny ImageNet} &
  \multirow{-3}{*}{Training} &
  \multirow{-2}{*}{Backdoor} &
  \cellcolor[HTML]{FFCCC9}Trigger &
  \cellcolor[HTML]{FFCCC9}66.14\% &
  \cellcolor[HTML]{FFCCC9}89.00\% $\in$ {[}87.44 , 90.79{]} &
  \cellcolor[HTML]{FFCCC9}0.660 $\in$ {[}0.608 , 0.716{]} &
  \cellcolor[HTML]{FFCCC9}0.904 $\in$ {[}0.891 , 0.920{]} &
  \cellcolor[HTML]{FFCCC9}52.15 \\ \hline
 &
   &
   &
  \multicolumn{2}{c}{Baseline} &
  94.35\% &
  81.42\% $\in$ {[}79.51 , 83.22{]} &
  0.646 $\in$ {[}0.617 , 0.679{]} &
  0.827 $\in$ {[}0.809 , 0.844{]} &
  82.23 \\
 &
   &
   &
  \multicolumn{2}{c}{Sponge-LBFGS} &
  - &
  80.94\% $\in$ {[}74.45 , 81.06{]} &
  0.638 $\in$ {[}0.522 , 0.640{]} &
  0.823 $\in$ {[}0.762 , 0.824{]} &
  81.75 \\
 &
   &
   &
  \multicolumn{2}{c}{Sponge-GA} &
  - &
  81.33\% $\in$ {[}81.27 , 81.40{]} &
  0.657 $\in$ {[}0.655 , 0.657{]} &
  0.826 $\in$ {[}0.825 , 0.827{]} &
  82.14 \\
 &
   &
   &
  \multicolumn{2}{c}{U-inputs} &
  - &
  80.32\% $\in$ {[}80.28 , 80.35{]} &
  0.640 $\in$ {[}0.639 , 0.640{]} &
  0.817 $\in$ {[}0.816 , 0.817{]} &
  81.12 \\
 &
   &
  \multirow{-5}{*}{Inference} &
  \multicolumn{2}{c}{Sparsity attack} &
  21.39\% &
  82.06\% $\in$ {[}80.31 , 83.96{]} &
  0.657 $\in$ {[}0.626 , 0.692{]} &
  0.821 $\in$ {[}0.803 , 0.840{]} &
  82.88 \\ \cline{3-10} 
 &
   &
   &
  \multicolumn{2}{c}{Poisoning} &
  94.27\% &
  82.21\% $\in$ {[}80.26 , 84.42{]} &
  0.659 $\in$ {[}0.626 , 0.699{]} &
  0.835 $\in$ {[}0.816 , 0.855{]} &
  83.03 \\
 &
   &
   &
   &
  \multicolumn{1}{l}{\cellcolor[HTML]{ECF4FF}Clean} &
  \cellcolor[HTML]{ECF4FF}93.15\% &
  \cellcolor[HTML]{ECF4FF}81.99\% $\in$ {[}80.10 , 84.33{]} &
  \cellcolor[HTML]{ECF4FF}0.655 $\in$ {[}0.623 , 0.696{]} &
  \cellcolor[HTML]{ECF4FF}0.833 $\in$ {[}0.815 , 0.855{]} &
  \cellcolor[HTML]{ECF4FF}82.81 \\
 &
  \multirow{-8}{*}{CIFAR-10} &
  \multirow{-3}{*}{Training} &
  \multirow{-2}{*}{Backdoor} &
  \multicolumn{1}{l}{\cellcolor[HTML]{FFCCC9}Trigger} &
  \cellcolor[HTML]{FFCCC9}91.97\% &
  \cellcolor[HTML]{FFCCC9}83.13\% $\in$ {[}81.57 , 84.77{]} &
  \cellcolor[HTML]{FFCCC9}0.676 $\in$ {[}0.649 , 0.707{]} &
  \cellcolor[HTML]{FFCCC9}0.843 $\in$ {[}0.829 , 0.859{]} &
  \cellcolor[HTML]{FFCCC9}83.96 \\ \cline{2-10} 
 &
   &
   &
  \multicolumn{2}{c}{Baseline} &
  72.66\% &
  80.73\% $\in$ {[}78.90 , 81.87{]} &
  0.628 $\in$ {[}0.595 , 0.655{]} &
  0.821 $\in$ {[}0.804 , 0.832{]} &
  82.34 \\
 &
   &
   &
  \multicolumn{2}{c}{Sponge-LBFGS} &
  - &
  77.49\% $\in$ {[}74.78 , 80.96{]} &
  0.575 $\in$ {[}0.526 , 0.639{]} &
  0.791 $\in$ {[}0.766 , 0.823{]} &
  79.04 \\
 &
   &
   &
  \multicolumn{2}{c}{Sponge-GA} &
  - &
  80.36\% $\in$ {[}80.35 , 80.38{]} &
  0.645 $\in$ {[}0.645 , 0.646{]} &
  0.817 $\in$ {[}0.817 , 0.817{]} &
  81.97 \\
 &
   &
   &
  \multicolumn{2}{c}{U-inputs} &
  - &
  81.17\% $\in$ {[}81.16 , 81.19{]} &
  0.656 $\in$ {[}0.656 , 0.656{]} &
  0.825 $\in$ {[}0.825 , 0.825{]} &
  82.79 \\
 &
   &
  \multirow{-5}{*}{Inference} &
  \multicolumn{2}{c}{Sparsity attack} &
  68.98\% &
  80.90\% $\in$ {[}79.00 , 82.16{]} &
  0.631 $\in$ {[}0.696 , 0.660{]} &
  0.809 $\in$ {[}0.790 , 0.821{]} &
  82.52 \\ \cline{3-10} 
 &
   &
   &
  \multicolumn{2}{c}{Poisoning} &
  70.94\% &
  82.64\% $\in$ {[}80.55 , 84.79{]} &
  0.663 $\in$ {[}0.621 , 0.713{]} &
  0.839 $\in$ {[}0.820 , 0.859{]} &
  84.29 \\
 &
   &
   &
   &
  \cellcolor[HTML]{ECF4FF}Clean &
  \cellcolor[HTML]{ECF4FF}69.39\% &
  \cellcolor[HTML]{ECF4FF}82.31\% $\in$ {[}80.38 , 84.05{]} &
  \cellcolor[HTML]{ECF4FF}0.657 $\in$ {[}0.618 , 0.698{]} &
  \cellcolor[HTML]{ECF4FF}0.836 $\in$ {[}0.818 , 0.852{]} &
  \cellcolor[HTML]{ECF4FF}83.96 \\
\multirow{-16}{*}{\rotatebox[origin=c]{90}{Mobilenet}} &
  \multirow{-8}{*}{Tiny ImageNet} &
  \multirow{-3}{*}{Training} &
  \multirow{-2}{*}{Backdoor} &
  \cellcolor[HTML]{FFCCC9}Trigger &
  \cellcolor[HTML]{FFCCC9}60.66\% &
  \cellcolor[HTML]{FFCCC9}83.23\% $\in$ {[}81.60 , 84.50{]} &
  \cellcolor[HTML]{FFCCC9}0.675 $\in$ {[}0.644 , 0.701{]} &
  \cellcolor[HTML]{FFCCC9}0.844 $\in$ {[}0.829 , 0.856{]} &
  \cellcolor[HTML]{FFCCC9}84.89 \\ \hline
\end{tabular}%
}
\end{table}

\subsection{Experimental Validation and Performance Analysis}
To further validate our analysis, we conducted a comparative experiment focusing on a selected subset of sparsity-based approaches. This subset was chosen due to the inherent differences in methodology and evaluation metrics across the broader range of attacks, which makes direct comparisons more challenging. Additionally, we concentrate on sparsity-based attacks because of their prominence in the current state-of-the-art, positioning them as a representative area for further investigation. In Table~\ref{tab:comparison}, we present the experimental results obtained using two widely adopted deep \ac{cnn} architectures for image classification: ResNet-18~\cite{resnet} and MobileNet-V2~\cite{mobilenets}. Specifically, we evaluate existing sparsity-based attacks from the literature, encompassing both inference- and training-stage attacks. Our experiments include inference-stage attacks such as Sponge attacks~\cite{shumailov2021sponge}, Uniform inputs~\cite{muller2024impact}, and sparsity attack~\cite{krithivasan2020sparsity}, as well as training-stage attacks, including the sponge poisoning attack~\cite{cina2022energy} and the energy backdoor attack~\cite{meftah2025energy}. These experiments were conducted on the CIFAR-10~\cite{cifar10} and Tiny ImageNet~\cite{tinyImageNet} datasets. The selection of these datasets is in line with the distinct characteristics of the evaluated attacks, which are relevant to both the inference and training stages. We report both sparsity-related metrics and actual energy estimations to provide a comprehensive evaluation of attack performance. Additionally, Figure~\ref{fig:box-plots} illustrates the distribution of energy ratios for each attack across the four model-dataset scenarios using box plots.

We observe that the poisoning attack~\cite{cina2022energy} can increase the model's overall energy consumption by up to 2mJ without significantly degrading its classification performance. However, a notable drawback of this approach is that it is designed to increases energy consumption across any input, thereby compromising its stealthiness. This characteristic is evident in Figure~\ref{fig:box-plots}, where the boxes corresponding to the poisoning attack shift upwards, indicating a consistent rise in the energy ratio across unaltered/clean images. Furthermore, poisoning attacks require access to the model's training phase, which, while potentially less realistic, is becoming increasingly feasible due to the growing adoption of \ac{mlaas} platforms and the widespread practice of downloading applications from the internet without thorough verification of their sources. One potential solution to mitigate this stealthiness limitation is the use of backdoor attacks, which are designed to maximize energy consumption for trigger samples while maintaining regular energy usage for clean inputs. Experimental results confirm the effectiveness of existing backdoor strategies in increasing energy consumption for trigger samples. However, these strategies require further refinement, as they may struggle to maintain low energy consumption for clean samples compared to the baseline model. This challenge is evident in the clean energy ratios observed in MobileNet-V2 trained on the Tiny ImageNet dataset. According to~\cite{meftah2025energy}, this can be attributed to the complexity of the objective function that includes two contrasting objectives (i.e, increasing energy consumption for trigger samples while decreasing it for clean samples) relative to the model's capacity, as well as the challenging nature of Tiny ImageNet. 

On the other hand, inference stage attacks face challenges in striking a balance between attack stealthiness and overall effectiveness. As highlighted in our analysis, while sparsity attack~\cite{krithivasan2020sparsity} maintain relatively good accuracy on adversarial inputs, their average energy consumption rates are closer to those of the baseline model, remaining lower than those of sponge and uniform attacks. An exception to this pattern is observed with the MobileNet model on the CIFAR-10 dataset, where the sparsity attack outperforms other inference stage attacks~\cite{shumailov2021sponge,muller2024impact} in terms of energy consumption. However, it fails to maintain good accuracy, with a decrease of up to 70\%, underscoring the trade-off between attack efficiency and stealthiness, at least with the approaches proposed in the literature. To improve the trade-off between stealth and effectiveness, we suggest employing more sophisticated attack adaptations, such as the \ac{pgd} attack~\cite{madry2017towards}, rather than directly optimizing the attack using its Lagrangian form. Additionally, we observe that the Sponge-\ac{lbfgs} attack may struggle to reach optimal solutions on average, as it sometimes generates adversarial images that consume less energy than clean images. This behavior is likely due to the sensitivity of the \ac{lbfgs} algorithm to initial conditions, a well-documented limitation of this optimizer~\cite{li2016learning}. On the other hand, Sponge-\ac{ga} appears to provide a more stable alternative, applicable in both black-box and white-box scenarios.

\section{Open Challenges and Future Works}
\label{sec:openchal}
\begin{figure}[t!]
    \centering
    \includegraphics[width=0.7\linewidth]{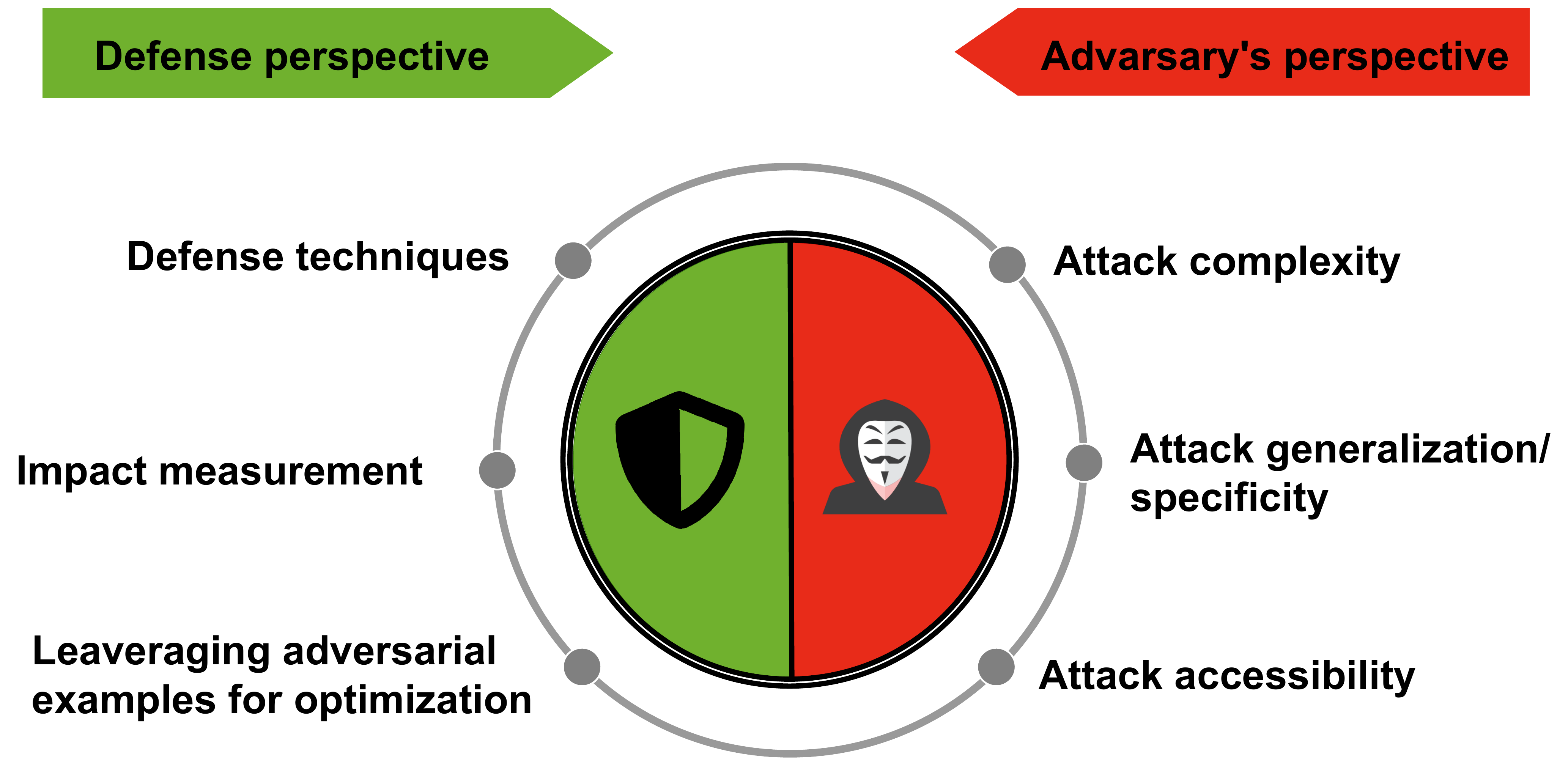}
    \caption{Open challenges from the defense and adversary's perspectives. }
    \label{fig:open_challenges}
\end{figure}

In this section, we outline open challenges in the context of sponge attacks and propose potential research directions to address these challenges. Fig.~\ref{fig:open_challenges} summarizes the discussed open challenges from both the adversary's and defender's perspectives.  \\\\
{\bf 1) Defense techniques.} Similar to traditional adversarial attacks, addressing vulnerabilities to energy-latency attacks is crucial for maintaining system robustness and security. In fact, traditional defenses designed for accuracy and privacy attacks are ineffective against energy-latency attacks~\cite{huang2024sponge,krithivasan2022efficiency}, due to the nature of these attacks that target the operational characteristics of a system, specifically aiming to disrupt its energy consumption patterns or increase the latency of operations in contrast to traditional attacks that target the model's overall accuracy or its confidentiality.
This implies that these attacks are very distinct in nature as they represent three distinct paradigms. While accuracy-based attacks involve optimizing the input to yield certain outputs, energy-latency attacks include disrupting the system by triggering its worst case energy consumption scenario.
Meanwhile, Krithivasan et \textit{al.}~\cite{krithivasan2020sparsity} explored potential defense techniques, such as activation thresholds, adversarial training, and compression, to counter attacks targeting sparsity-based accelerators. However, their experiments revealed that these defenses did not substantially mitigate such attacks without adversely impacting model accuracy. Currently, the only proposed defense strategy is {\bf DefQ} by Qiu et \textit{al.}~\cite{qiu2021defq}, which employs a pre-processing quantization function in front-end sensors to mitigate adversarial perturbations. This is achieved through a quantization table derived from a statistical analysis of {\bf DeepSloth} adversarial samples in the \ac{dct} frequency domain. While effective against {\bf DeepSloth}, its efficacy against other energy-latency attacks remains uncertain, highlighting the need for more generalized defense mechanisms.

Therefore, investigating more robust defense techniques is a promising avenue for further research. For \ac{cv} applications, pre-processing mechanisms present an intriguing direction. Various input transformations, like noise injection, smoothing, geometric transformations, and quantization, can reduce adversarial perturbations in images, while reconstruction models such as diffusion models, generative models, and auto-encoders can potentially remove adversarial noise. For \ac{nlp} applications, pre-processing steps that filter characters based on their Unicode encoding may help detect potential homoglyphs in user input. Additionally, the ratio of tokens categorized as unknown by the tokenizer could indicate whether the input is clean or adversarial.

While these suggestions offer valuable research directions, it is crucial to consider algorithmic complexity and energy consumption during the development of defense and detection mechanisms. For these mechanisms to be effective, their computational and energy costs must be lower than the potential costs induced by energy-latency attacks. One promising option is to integrate these objectives into existing pre-processing defense functions within the prediction pipeline or develop optimized multi-objective pre-processing strategies that prioritize simultaneous defense against multiple threats over threat-specific defenses. \\\\
\textbf{2) Impact of measurement.} Many current strategies assume a white-box setting, where the attacker has full knowledge of the target model. Designing attacks that do not rely heavily on this knowledge is important for real-world scenarios. 

Existing metrics provide good indicators of attack performance in terms of latency, battery drain, and energy consumption.
However, the majority of metrics used in the literature are application-specific, which hinders a unified comparison of existing attacks based on tangible performance such as energy consumption, greenhouse gas emission or latency.
Moreover, further investigation into the long-term economic and environmental consequences of these attacks is needed.  \\\\
\textbf{3) Leveraging adversarial examples for optimization.} Muller \textit{et al.}~\cite{muller2024impact}  showed the potential of using adversarial examples to optimize \ac{dnn} sparsity. Exploring this approach to improve model efficiency is a promising research direction. \\\\
{\bf 4) Attack complexity.} Designing attacks often involves optimizing adversarial examples to meet specific goals. Although some strategies allow optimization on the victim's server using queries~\cite{shumailov2021sponge},  most of the optimization strategies adopted in the literature~\cite{haque2020ilfo,hong2020panda,pan2022gradauto,chen2022nmtsloth} shift the burden to the attacker. This optimization can induce computational and energy costs to the adversary. Simplifying this process and the investigation of more computationally efficient attack strategies is crucial, especially in real-time attack scenarios. \\\\
{\bf 5) Attack generalization/specificity.}  Finding a balance between general and application-specific attacks is essential to achieve the desired impact with minimal expertise. In fact, general attacks can be broadly applicable to different architectures in different contexts, but their success rate may be potentially low as these attacks are designed to exploit common vulnerabilities across a wide range of systems. Moreover, these attacks are more likely to be detected and mitigated by common security measures, rendering them even less effective. On the other hand, application-specific attacks are tailored to exploit unique vulnerabilities within a given model making them more effective in both terms of success rate and non-detectability compared to general attacks, but the development of such attacks can require a higher level of expertise to identify and exploit these specific vulnerabilities and can considerably narrow the range of targeted models.
Developing attacks that are both effective and broadly applicable remains a challenge if not impossible, but finding a good balance between general and application-specific attacks can be an interesting research direction.  \\\\
\textbf{6) Attack accessibility.} Many current strategies assume a white-box setting, where the attacker has full knowledge of the target model.
While it is logical for initial research to focus on white-box contexts to uncover the maximum number of vulnerabilities and have a better understanding of these attacks, it is equally important to design attacks that rely less on this knowledge for real-world applicability in the long term.

\section{CONCLUSION}
\label{sec:conclusion}
In this paper, we provided a comprehensive overview of energy-latency attacks against \acp{dnn} proposed in the literature. We categorized these attacks using the established taxonomy for traditional adversarial attacks and presented relevant background information for each target application before delving into the details of each attack strategy. We also discussed various metrics used to assess attack success and provided a comparative analysis of these works. Finally, we presented existing defense strategies against these attacks and highlighted open challenges and future research directions in this field.

\begin{acks}
This work is fully funded by R\'egion Bretagne (Brittany region), France, CREACH Labs and Direction G\'en\'erale de l'Armement (DGA).

\end{acks}

\bibliographystyle{ACM-Reference-Format}
\bibliography{acmart}

\end{document}